\documentclass[]{emulateapj}

\usepackage{amsmath}
\usepackage{longtable}
\usepackage{multirow}

\newcommand{\teff}{${T}_{\rm eff}$}
\newcommand{\feh}{[Fe/H]}
\newcommand{\logg}{\emph{log}g}
\newcommand{\U}{$\langle U \rangle$}
\newcommand{\V}{$\langle V\rangle$}
\newcommand{\W}{$\langle W\rangle$}
\newcommand {\vrq}{\overline{\upsilon^2_{\!R}}}
\newcommand {\va}{\upsilon_{\mathrm{a}}}

\newcommand {\Usun}{{U_{\!\odot}}}
\newcommand {\Wsun}{{W_{\!\odot}}}
\newcommand {\Vsun}{{V_{\!\odot}}}
\newcommand{\revise}[1]{\textbf{#1}}

\def\bolsig{\mbox{\boldmath$\sigma$}}

\def\kpc{\,\mathrm{kpc}}
\def\kms{\,\mathrm{km\,s}^{-1}}

\def\bolv{\mathbf{v}}

\def\pc{\,\mathrm{pc}}

\def\pc{\,\mathrm{pc}}

\def\bolSig{\mbox{\boldmath$\Sigma$}}
\def\bolA{\mbox{\boldmath $A$}}
\def\bolB{\mbox{\boldmath $B$}}
\def\bolD{\mbox{\boldmath $D$}}
\def\bolV{\mbox{\boldmath $V$}}
\def\bolLamb{\mbox{\boldmath $\Lambda$}}

\begin{document}


\title{The stellar kinematics in the solar neighborhood from LAMOST data}


\author{Hai-Jun Tian\altaffilmark{1,2}, Chao Liu\footnote{corresponding author}\altaffilmark{2}, Jeffrey L. Carlin\altaffilmark{3, 4},Yong-Heng Zhao\altaffilmark{2}, Xue-Lei Chen\altaffilmark{2}, Yue Wu\altaffilmark{2}, Guang-Wei Li\altaffilmark{2}, Yong-Hui Hou\altaffilmark{5}, Yong Zhang\altaffilmark{5}}

\affil{
1. China Three Gorges University, Yichang, 443002\\
2. Key Laboratory of Optical Astronomy, National Astronomical Observatories, Chinese Academy of Sciences, Beijing 100012\\
3. Department of Physics, Applied Physics and Astronomy, Rensselaer Polytechnic Institute, 110 8th Street, Troy, NY 12180, USA\\
4. Department of Physics and Astronomy, Earlham College, 801 National Road West, Richmond, IN 47374, USA\\
5. Nanjing Institute of Astronomical Optics \& Technology, National Astronomical Observatories, Chinese Academy of Sciences, Nanjing 210042, China}

\begin{abstract}
We use about 200,000 FGK type main-sequence stars from the LAMOST DR1 data to map the local stellar kinematics. With the velocity de-projection technique, we are able to derive the averaged 3 dimensional velocity and velocity ellipsoids using only the line-of-sight velocity for the stars with various effective temperatures within $100 < |z| < 500$ pc. Using the mean velocities of the cool stars, we derive the solar motion of ($\Usun$, $\Vsun$, $\Wsun$)=(9.58$\pm2.39$, 10.52$\pm1.96$, 7.01$\pm1.67$)\,$\kms$ with respect to the local standard of rest. Moreover, we find that the stars with \teff$>6000$\,K show a net asymmetric motion of $\sim3$\,$\kms$ in \W\ compared to the stars with \teff$<6000$\,K. And their azimuthal velocity increases when $|z|$ increases. This peculiar motion in the warmer stars is likely because they are young and not completely relaxed, although other reasons\revise{, such as the resonance induced by the central rotating bar or the spiral structures, and the perturbation of the merging dwarf galaxies, can not be} ruled out. The derived velocity dispersions and cross terms for the data are approximately consistent with previous studies. We also find that the vertical gradients of $\sigma_{U}$ and $\sigma_V$ are larger than that of $\sigma_W$ . And the vertical gradient of $\sigma_U$ shows clear correlation with \teff , while the other two do not. Finally, our sample shows vertex deviation of about 11$^\circ$, at $300 < |z| < 500$\,pc, but roughly zero at $100 < |z| < 300$\,pc.

\end{abstract}


\maketitle

\section{Introduction}\label{sec_intro}

The velocity distribution of the stars in the solar neighborhood plays a key role in understanding the global structure, dynamical features, and the evolution of the Milky Way. Although it is often approximated with a multi-dimensional Gaussian profile, the velocity distribution of the stars in the solar neighborhood is actually very complicated. The mean value of the velocity distribution should be around zero, given that the Galactic disk is in static state. However, observations have found many substructures \citep[][]{dehnen1998,zhao2009,siebert11, A12,xia14}, which may be associated with the perturbation of the Galactic bar and spiral arms, or belong to old tidal debris of disrupted clusters or dwarf galaxies \citep[][]{dehnen2000,fux2001,famaey2005,antoja2011}, in the velocity distribution. These substructures may shift the mean velocity slightly away from zero by a few $\kms$ .

Recently, \citet[][]{widrow2012} found that the stellar number density is not symmetric with respect to the Galactic mid-plane, implying a vertical wave in the stellar disk. \citet{gomez2013} inferred that the merging event of the Sagittarius dwarf galaxy can induce such density waves. The vertical asymmetry in stellar count may be associated with vertical asymmetry in velocity. Indeed, \citet[][]{carlin2013} found from the LAMOST DR1 data that not only radial velocity but also the vertical velocity of the nearby stars are not symmetric. These new observational evidences challenge the current dynamical models of the disk. At least some of the nearby disk stars are not in equilibrium, although the majority must be in the static state so that the Galactic disk can survive over billions of years. More investigation is required to figure out how many and which types of stars contribute to the asymmetric motion, which is our main motivation of this work.

The velocity distribution can be characterized by the velocity ellipsoid, which reflects the mass distribution and evolution of the Milky Way, assuming that most of the detected stars are in equilibrium. The earliest study of the stellar velocity ellipsoid was done by \citet{Schwarzschild1908}. From then on, many works have found that the age of stars is correlated with the velocity distribution. Specifically, older stars show larger velocity dispersion, and vice versa \citep[][etc.]{parenago1950, Roman1950, Roman1952, DB98, quillen2001,N04, holmberg2007}. This is usually thought to be because scattering of the disk stars increases over time. For the stars younger than $\sim$8 Gyr, the scattering is most likely due to encounters with substructures, e.g., giant molecular clouds, in the disk \citep{holmberg2007}. The age--velocity dispersion relation (AVR) reflects the evolution history of the Galactic disk.

In an ideal axisymmetric disk, the velocity ellipsoid near the disk mid-plane should be aligned with the cylindrical coordinates. However, since the Galactic disk contains a rotational central bar and a number of spiral arms, the velocity ellipsoid will not match this ideal case. On the contrary, it deviates from the cylindrical coordinates near the Galactic mid-plane. \citet[][hereafter DB98]{DB98} reported that the vertex deviation of the velocity ellipsoid measured from \emph{Hipparcos} proper motion data is $\sim10^\circ$. \citet[][hereafter S12]{smith2012} confirmed that the vertex deviation of the metal-rich stars in the SDSS sample is consistent with DB98. Moreover, the authors also showed that the tilt angle relative to the Galactic mid-plane is about $-10$ to $-15\degr$ for the velocity ellipsoid at $z>0.5\kpc$. Recently, lots of new works have been done with the data from large stellar spectroscopic surveys, e.g., \citet{B14} and \citet{Sharma14} based on RAVE~\citep{rave}, \citet{bovy14} based on APOGEE~\citep{ahn14}, and \citet{racio14} based on Gaia-ESO survey~\citep{gaiaeso}.

Since 2011, the LAMOST survey \citep[][]{Cui2012, Zhao2012, Deng2012} has been operating, and about 1 million stellar spectra with stellar parameters have been released in the DR1 catalog. A large fraction of these stars are located within 1$\kpc$ around the Sun, providing a vast resource to reveal details of the local stellar velocity distribution and give constraints on the dynamical structure and evolution of the Galactic disk.

In this paper, we use the FGK type main-sequence stars selected from the LAMOST DR1 catalog to study the local velocity distribution. We use the effective temperature of these stars as a proxy for age and investigate the variation of the velocity distribution, including the mean velocity and the velocity ellipsoid, as a function of the effective temperature and height above/below the Galactic mid-plane.

The paper is organized as follows. In Section 2, we describe how we select the samples. In Section 3, we introduce a velocity de-projection method to reconstruct the velocity and velocity ellipsoid in three dimensions from only the line-of-sight velocity. The method is then validated with GCS data and a mock catalog. The systematic biases due to the uneven spatial sampling and the spatial variation of the velocity ellipsoid are then calibrated. The mean velocities and velocity ellipsoids in various effective temperature bins at different height is then derived and discussed in Sections 4 and 5. Finally, brief conclusions are drawn in Section 6.

\section{The Data}\label{sec_data}
\subsection{The LAMOST Survey}\label{sec_lamost}
LAMOST is a quasi-meridian reflecting Schmidt telescope with an effective aperture of about 4m and a field of view of 5$\degr$, operated by the National Astronomical Observatories, Chinese Academy of Sciences. It is a powerful instrument to survey the sky with the capability of recording 4000 spectra simultaneously \citep{Cui2012, Zhao2012}. In its five-year survey plan, it will obtain a few millions stellar spectra in about 20,000 square degrees in the northern hemisphere \citep{Deng2012}.

\begin{figure}[!t]
\includegraphics[width=0.5\textwidth, trim=0.5cm 0.5cm 0.5cm 0.5cm, clip]{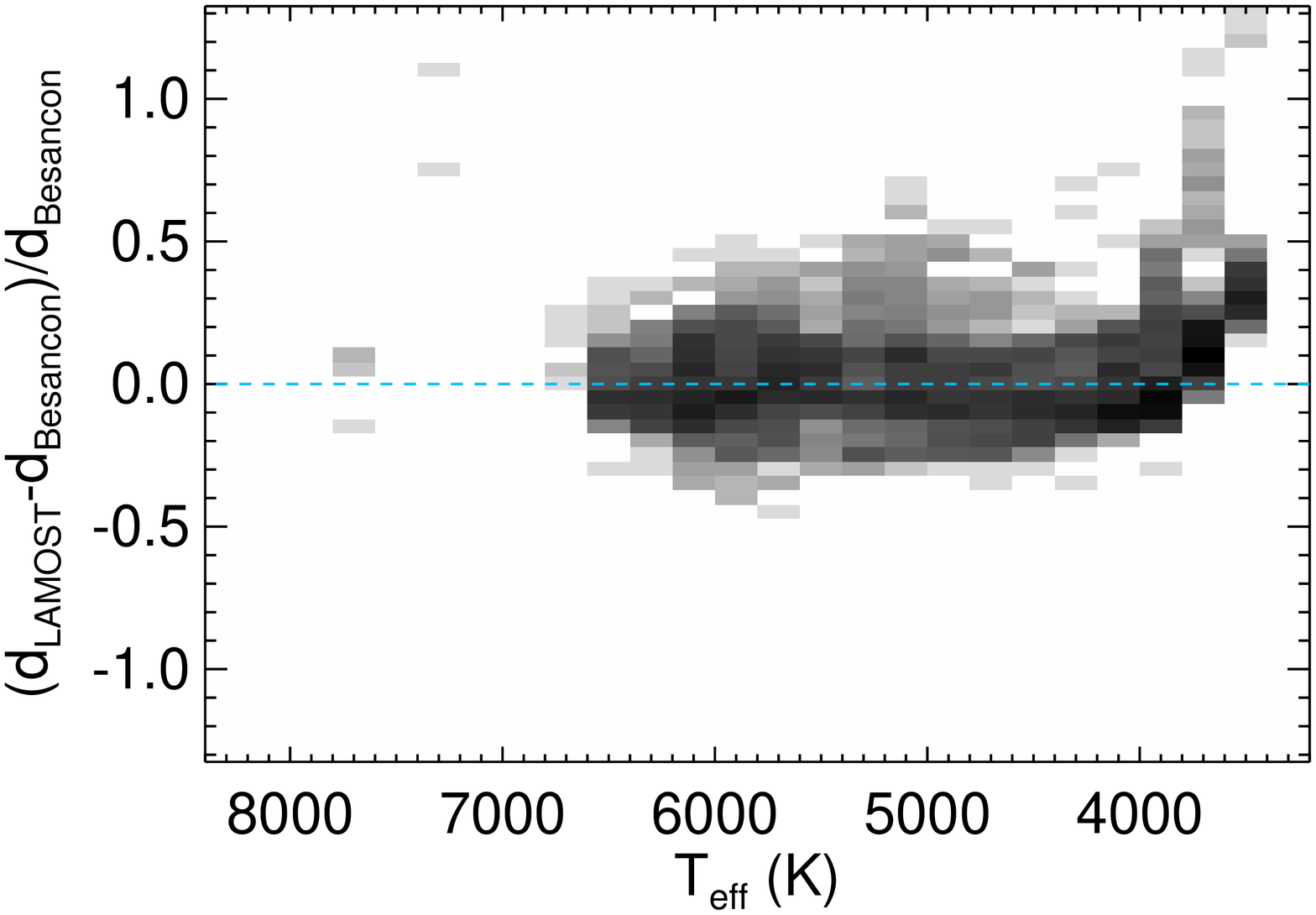}
\caption{\revise{Fractional distance residuals resulting from running the \citet{carlin15} distance code on a realizations of the Besancon model in a field at $(l,b) = (180\degr,60\degr)$.}}\label{fig:distresid}
\end{figure}

The LAMOST Survey has internally delivered the first data release (DR1), which contains 2,204,860 spectra with a resolution of $\sim$1800 covering a wavelength range of $3800 \lesssim \lambda \lesssim 9100 \AA$. The catalog contains 1,085,404 stellar spectra with estimated stellar atmospheric parameters as well as line-of-sight velocities. The distances to stars with estimated effective temperature, surface gravity, and metallicity are determined from isochrone fitting by ~\citet{carlin15}. They developed two techniques, chi-square and Bayesian to estimate the absolute magnitude. The chi-square technique finds the best fit absolute magnitude which reaches the minimum of the chi-square \revise{between the measured \teff\, log$g$, [Fe/H] and the isochrone ones}. The Bayesian technique, on the other hand, considers both the selection effect and the different stellar populations along different line of sight. They verified that the accuracies of the distance estimates derived from both techniques are $\sim20$\%. They also found that their method may underestimate the distance of the giant stars by about 20\%, but for main-sequence stars \revise{with $4000<$\teff$<7000$\,K, no correlation is found with \teff, as presented in Figure~\ref{fig:distresid}}. Compared with the \emph{Hipparcos} data, they obtained fairly good distance estimates with scatter of only $\sim17$\%. Considering that only the FGK main-sequence stars are used in this work, the accuracy of the distance is sufficient for the studies of the local stellar kinematics within about 1\,kpc. For such nearby stars, not significant difference is found between the two techniques. Therefore, we adopt the distance estimates from the chi-square technique in this work.

\subsection{The sample selection}\label{sec_selectiondata}
\begin{figure}
\begin{center}
\includegraphics[height=0.51\columnwidth]{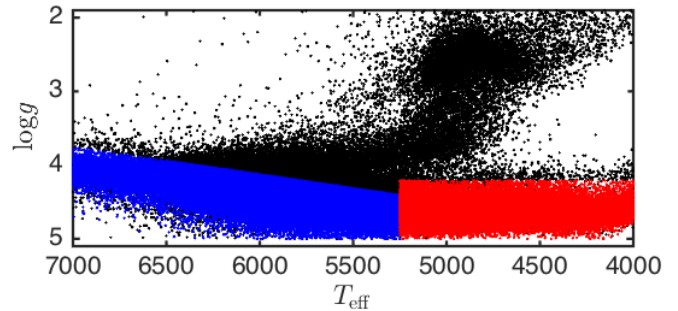}
\end{center}
\caption{The distribution of the LAMOST DR1 stars in \teff\ vs. log$g$ plane. The black points are all stars in the LAMOST DR1 with signal to noise ratio of $SNR>10$ in $g$ band . The blue and red points are the main-sequence stars selected for this work. }\label{fig_teff_logg_a}
\end{figure}

\begin{figure}
\begin{center}
\includegraphics[height=1.0\columnwidth]{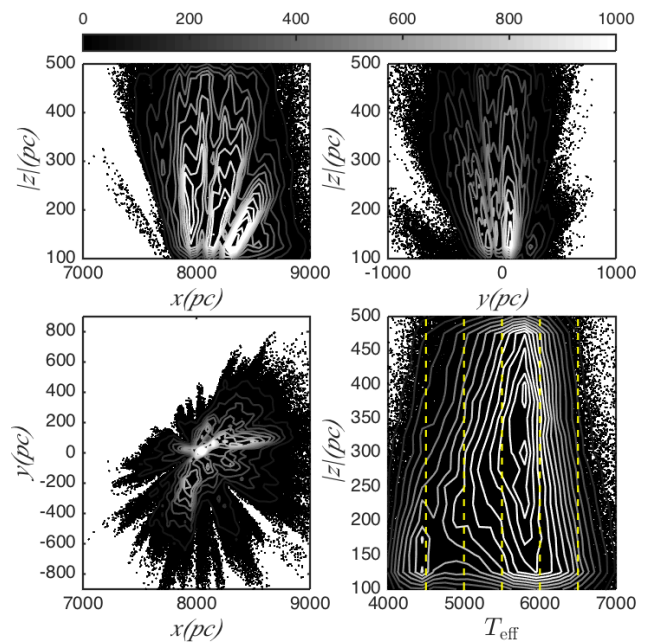}
\end{center}
\caption{The distribution of the sample in Galactocentric Cartesian coordinates. The top-left (x-$|z|$), top-right (y-$|z|$), and bottom-left (x-y) panels shows the projected distributions, respectively. The Sun is located at $(x,y,z)$ = (8.0, 0, 0) $\kpc$. The bottom-right panel shows the stellar distribution in \teff\ vs. $|z|$ plane.  The five vertical dashed lines divide \teff\ into six bins, in which the velocities and their dispersions are determined in this work. \revise{The contours represent the number of stars with different levels.}}\label{density_distri}
\end{figure}

\begin{figure}
\begin{center}
\includegraphics[height=0.47\columnwidth]{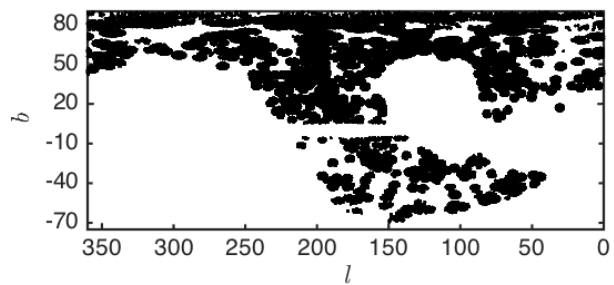}
\end{center}
\caption{The distribution of the sample in $l$ vs. $b$ plane.}\label{data_distr}
\end{figure}

We select nearby FGK type main-sequence stars to investigate the kinematics of the solar neighborhood. In order to select main-sequence stars we define the selection criteria separately for stars with \teff\ larger and smaller than 5250\,K. For the stars with \teff\ higher than 5250\,K (displayed by the blue dots in Figure \ref{fig_teff_logg_a}), the selection criteria are as follows:
\begin{itemize}
    \item 5250$<$\teff$<$7000 K,
    \item $\log{g} >$ (3.75-4.5)/(7000-5000)*(\teff-5000)+4.5,
   \item (-\teff/500 + 14.1)$<M_{J}<$(-\teff/550 + 15.0),
\end{itemize}
where $M_J$ is the estimated absolute magnitude in $J$ band from Carlin et al. For stars with \teff$<$5250\,K (presented by the red points in Figure \ref{fig_teff_logg_a}), the criteria are as follows:
\begin{itemize}
    \item 4000$<$\teff$<$5250 K,
    \item $\log{g} >$ 4.2,
   \item (-\teff/700 + 11.0)$<M_{J}<$(-\teff/750 + 12.0).
\end{itemize}
The criteria on $\log{g}$ and $M_J$ are empirically set to ensure that the selected stars are in the main-sequence. The criteria for higher \teff\ remove the possible sub-giant branch stars so that the rest of the warm main-sequence stars may better trace the young populations. The criteria for lower \teff\ exclude the giant stars.
We further select stars whose spectra have signal-to-noise ratio higher than 10 in the $g$ band, which is the most sensitive portion of the spectrum to the stellar parameters. We also select the stars within
\begin{itemize}
    \item 100$<|z|<$500\,pc,
    \item 7$<R_{GC}<$9\,kpc,
\end{itemize}
adopting the position of the Sun to be $(X,Y,Z)$ = (8, 0, 0) kpc. The stars within 100$\pc$ are not well observed in the LAMOST survey because most of these stars are too bright. Therefore, we only select the stars with $|z|>100$\,pc. Finally, a total of 209,316 FGK type main-sequence stars are selected after applying all above criteria. It is worthy to point out that this sample is significantly extended compare to the previous works \citep[DB98; S12;][etc.]{N04,siebert08,BvW14}. Therefore, the statistics based on such a large sample may be substantially improved and the sample containing a wide range of effective temperatures allows us to map the detailed kinematics for variant spectral types. Figure \ref{density_distri} shows the distribution of the sample in the Cartesian coordinate system with respect to the Galactic center (the top-left, top-right and bottom-left panels) and the \teff\ vs. $|z|$ plane (the bottom-right panel). The five vertical dashed lines in the bottom-right panel divide the effective temperature into six bins, in which the stellar kinematics are determined separately in the later sections. Figure \ref{data_distr} presents the distribution of the sample in \revise{$l$ vs. $b$} plane. From these figures, it is clear that the stars are not uniformly distributed in the sky. This leads to a distortion of the measured velocity ellipsoid, which is discussed in the Section \ref{sec_mock}. Figure \ref{fig_snrg} displays that most of the selected stars are with the signal to noise ratio in $g$ band larger than 10.

\begin{figure}
\begin{center}
\includegraphics[height=0.5\columnwidth]{hist_snrg.eps}
\end{center}
\caption{The distribution of the signal to noise ratio in $g$ band of the sample.}\label{fig_snrg}
\end{figure}

\subsection{The radial velocity}\label{sec_radialvel}
\begin{figure*}[htbp]
\centering
\begin{minipage}{18cm}
\centering
\includegraphics[scale=0.55]{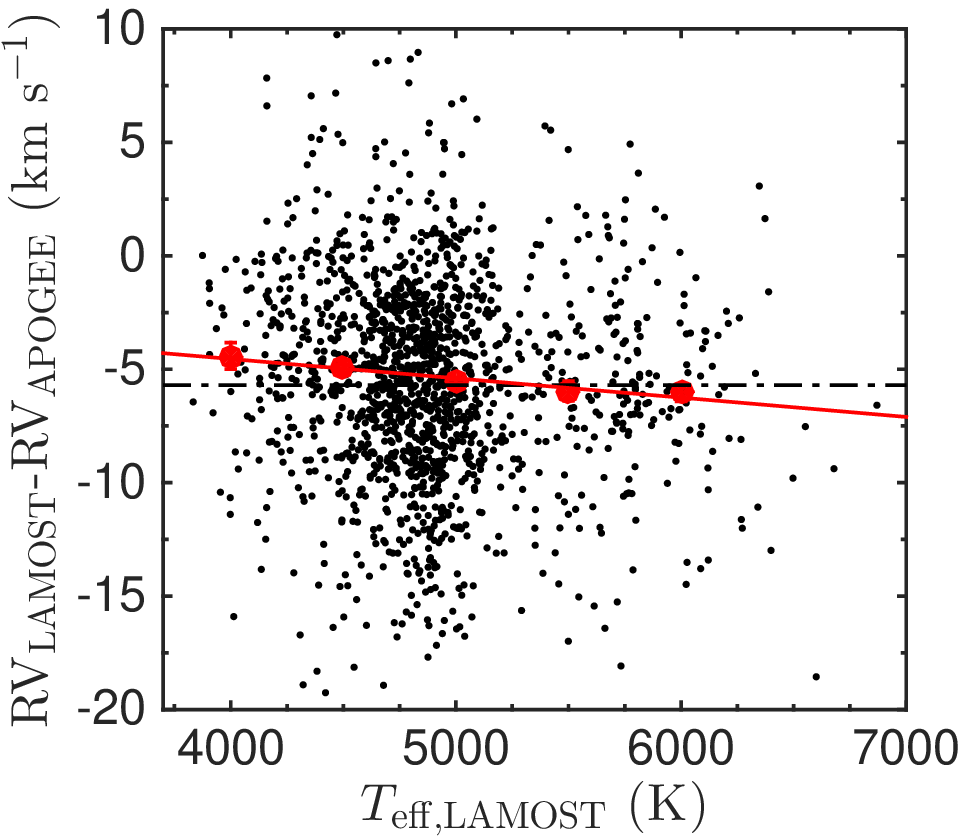}
\includegraphics[scale=0.55]{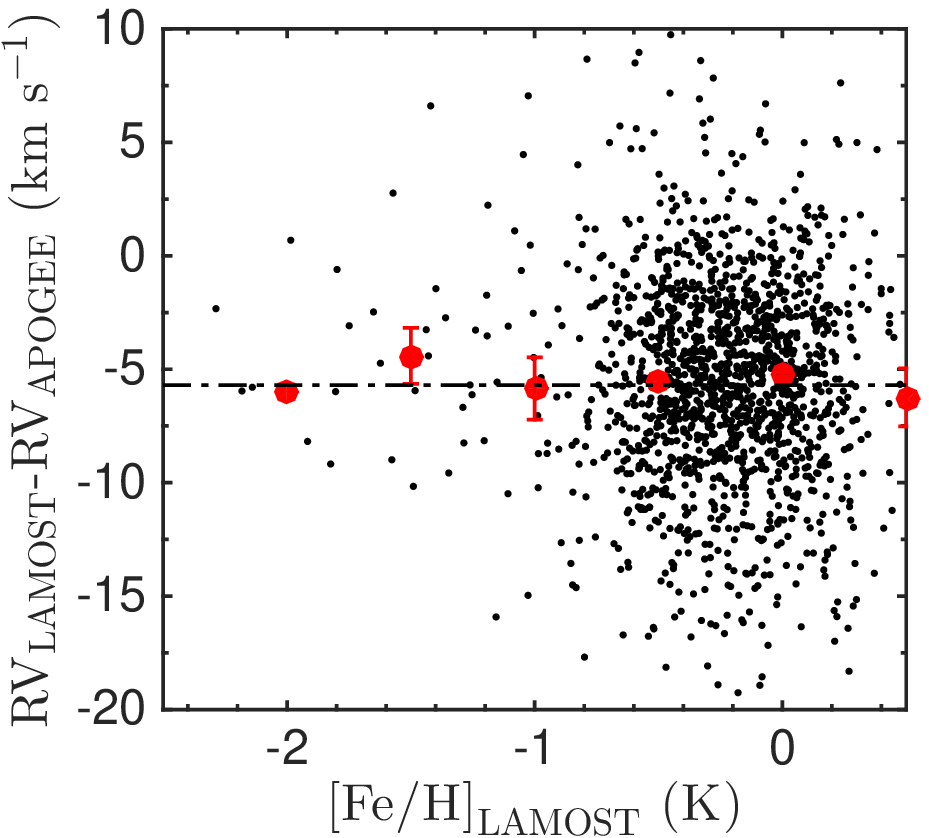}
\includegraphics[scale=0.55]{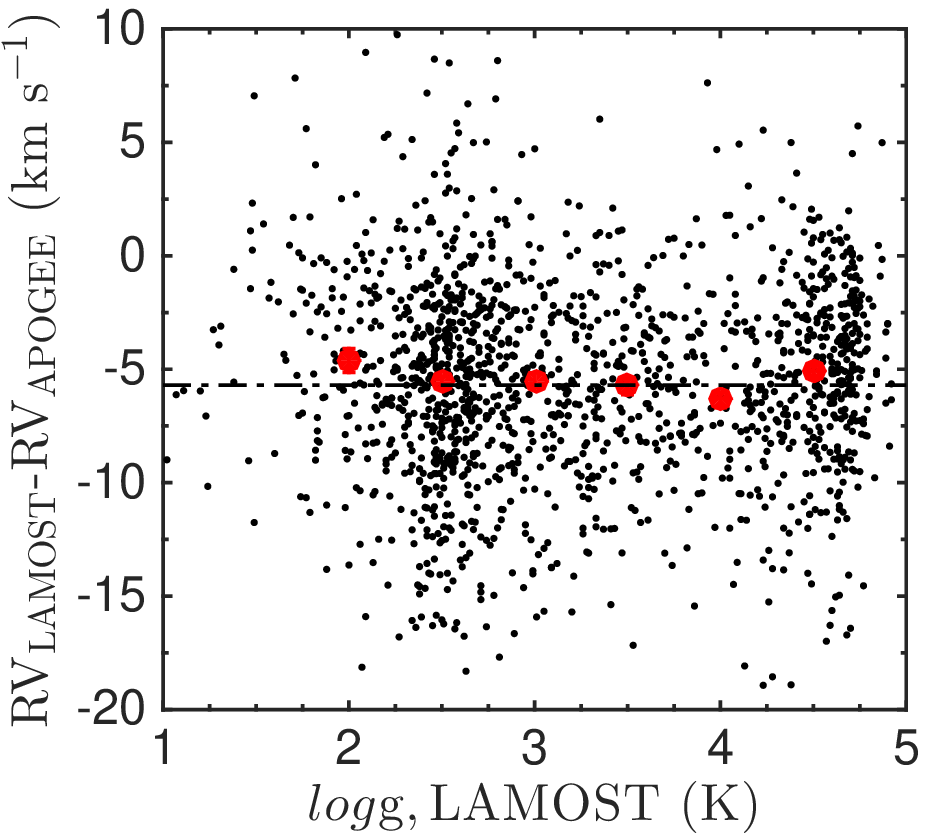}
\end{minipage}
\caption{\revise{Left panel: The offset of the radial velocity between LAMOST and APOGEE as the function of the LAMOST \teff. The black dots are the individual stars and the red filled circles are the median values in each \teff\ bin.} \revise{The red line indicates the best linear fit of the median points. The black dotted-dash line marks the location of the average offset of 5.7 $\kms$.} \revise{Middle panel: The offset of the radial velocity as a function of the LAMOST \feh. The symbols are similar as the left panel. No clear correlation is found in this plot. } \revise{Right panel: The offset of the radial velocity as a function of the LAMOST \logg. The symbols are similar as the left panel.}}\label{fig_lm_apo}
\end{figure*}

The LAMOST catalog provides the radial velocity of each star with uncertainty of about 5\,$\kms$ \citep{gao14}. In order to investigate the systematics of the radial velocities, we cross identify the LAMOST data with the APOGEE data released in SDSS DR10 \citep{ahn14} and find $\sim$1000 common stars with good parameter estimates and velocity scatter smaller than 0.3\,$\kms$ (which removes most multiple or variable stars) from APOGEE and signal-to-noise ratio higher than 20 in the LAMOST spectra. The radial velocity derived from the LAMOST pipeline is slower by 5.7\,$\kms$ compared with APOGEE, as shown in Figure~\ref{fig_lm_apo}. The reason for this offset is unclear and is worthy of further investigation in future works. In this work, we simply add an additional 5.7\,$\kms$ to the LAMOST derived radial velocity to match the other survey data. \revise{It is noted that the offset is weakly anti-correlated with \teff, but no significant correlation with \feh\ and \logg. We further investigate whether the weak anti-correlation with \teff\ changes our result in section~\ref{sect:sysbias}}. 

%
%
%

\section{The method}\label{sec_method}
For most of the LAMOST data, the proper motions can be found from either UCAC4~\citep{ucac4} or PPMXL~\citep{ppmxl} catalog. However, because the purpose of this work is to derive the velocity ellipsoids, it requires accurate measurement of the uncertainties of the proper motions as well as the values themselves. Unfortunately, after \revise{tentatively measuring the three dimensional velocities and their ellipsoids with the combination of the proper motions and the radial velocity}, we find that the errors of these proper motions are too large to derive the reliable velocity ellipsoids. Therefore, we turn to measure the three dimensional velocity and its ellipsoid from only the one dimensional line-of-sight velocities of the LAMOST sample spanning a large area of the sky. In this section, we first describe the velocity de-projection method and then discuss the validation and calibration of the approach.

\subsection{The de-projection method}\label{sec_deproj}
Consider a star with the line-of-sight unit vector ${\mathbf{r}}$ and heliocentric velocity of ${\mathbf v}$. The observed line-of-sight velocity $v_{los}$ can be written as
\begin{equation}\label{eq:vproj}
v_{los}={\mathbf{r}}\cdot{\mathbf v}.
\end{equation}
${\mathbf r}$ is determined by the Galactic coordinates $l$ and $b$ and the three components of ${\mathbf v}$ in the Galactocentric Cartesian coordinates are $U$, $V$, and $W$, respectively. Then Equation~(\ref{eq:vproj}) can be expanded as
\begin{equation}\label{v_los}
 v_{los}= U{\rm cos}l{\rm cos}b+V{\rm sin}l{\rm cos}b+ W{\rm sin}b.
\end{equation}
Given a group of stars at similar distance along the line of sight ${\mathbf r}$, the velocity dispersion projected onto the line of sight should be
\begin{equation}\label{eq:sproj}
{\sigma}^2_{los}={\sum_{i=1}^{N}{({\mathbf r_i}\cdot({\mathbf v_i}-\overline{{\mathbf v}}))^2}\over{N}},
\end{equation}
where N is the number of the stars. This can be specified as
\begin{equation} \label{sigma_los}
\begin{split}
{{\sigma}^{2}_{los}} &= \sigma_{U}^{2}{\rm cos}^2l{\rm cos}^2b + \sigma_{V}^{2}{\rm sin}^2l {\rm cos}^2b  + \sigma_{W}^{2}{\rm sin}^{2}b
 \\&+ \sigma_{UV}^{2}{\rm sin}2l {\rm cos}^{2}b + \sigma_{UW}^{2}{\rm cos}l {\rm sin}2b
\\&+ \sigma_{VW}^{2} {\rm sin}l {\rm sin}2b,
\end{split}
\end{equation}
where the dispersions $\sigma_{U}$, $\sigma_{V}$ and $\sigma_{W}$ specify the size and shape of the velocity ellipsoids, and the cross terms $\sigma_{UV}$, $\sigma_{UW}$ and $\sigma_{VW}$ determine the orientation of the velocity ellipsoids.

\subsection{The likelihood}\label{sec_lklhd}
According to Equations \ref{v_los} and \ref{sigma_los}, one is able to reconstruct the 3-D mean velocity and velocity ellipsoid at a given spatial position from the line-of-sight velocities of a group of stars located at the same position. To do this we construct a likelihood as the following
\begin{equation} \label{likelihood}
L = \prod_{k = 1}^{N}\bigg\{\frac{1}{\sqrt{{\sigma}_{model,k}^2 + \epsilon_{k}^2}} \exp\left[-\frac{(v_{los,k} - {v}_{model,k})^2}{2({\sigma}_{model, k}^2 + \epsilon_{k}^2)}\right]\bigg\},
\end{equation}
where $v_{los,k}$ is the measured line-of-sight velocity of $k$th star, $\epsilon_{los,k}$ \revise{is} the measured error of the line-of-sight velocity, ${v}_{model, k}$ and ${\sigma}_{model, k}$ \revise{are} the predicted mean line-of-sight velocity and the projected velocity dispersion along the line-of-sight of the $k$th star, respectively. The following 9 quantities, $U$, $V$, $W$, $\sigma_{U}$, $\sigma_{V}$, $\sigma_{W}$, $\sigma_{UV}$, $\sigma_{UW}$, and $\sigma_{VW}$ are free parameters in Equation~(\ref{likelihood}). The best fit values for them can be found by maximizing the likelihood $L$.

We use Markov chain Monte Carlo (MCMC) simulation to estimate the posterior probabilities of the velocities and the velocity ellipsoids. The priors of the three mean velocities, $U$, $V$, and $W$, are evenly distributed in all real values, while the priors of the velocity ellipsoids are evenly distributed in the range of (0, $+\infty$). In this work, we  run the MCMC simulation with $\bf emcee$, which implements the affine-invariant ensemble sampler \citep{Goodman2010}.

\subsection{Validation of the method}\label{sec_vali}
The velocity de-projection technique has been applied by DB98, who used the proper motions of \emph{Hipparcos} to estimate the velocity and dispersions of stars with median distance of $\sim$70\,pc around the Sun. Later, \citet{F09} applied the similar method to the SDSS data, which extends to a few hundred parsecs away from the Sun. However, \citet{MB09} then pointed out that the de-projection technique may produce systematic bias in the cross terms, especially in $\sigma_{UW}$, if $\sigma$ is a function of $R$ and $z$. Because our method is very similar to that DB98 used and our data is located between 100 and 500\,pc in $|z|$, we may also suffer from similar biases. Furthermore, unlike the \emph{Hipparcos} data, the highly uneven spatial distribution of the LAMOST data may lead to another systematic bias. Therefore, before applying this method to the LAMOST data, we need to carefully validate it and understand these issues.

\subsubsection{Validation with GCS data}\label{sect:gcsvalid}
Our first validation uses data from the Geneva-Copenhagen Survey \citep[GCS;][]{N04}, which is a volume-complete set of F- and G-type stars within $\sim$40\,pc, and provides three dimensional velocity components for each individual star. This dataset does not extend to a large volume, hence it may not be severely affected by the systematic bias explained by \citet{MB09}. Thus we expect that the de-projection method may give estimates close to the true values.

A sub-sample of 3712 dwarf stars for the validation test is selected from GCS data with 5500 K$<$\teff$<$6000 K, $|U|<200 \kms$, $|V|<400 \kms$, and $|W|<200 \kms$. Outliers beyond 4$\sigma$ in the line-of-sight velocity distribution are removed. Only the line-of-sight velocity of these data is used to derive their mean velocities and velocity ellipsoid according to Equation~(\ref{likelihood}) via MCMC simulation. In order to reduce the effect of the uneven distribution on the sky, we separate the sky into small equal-area bins and randomly select roughly equal numbers of stars in each bin so that the samples are approximately evenly distributed on the whole sky. We repeat the arbitrary draw of the data in equal-area bins 40 times and estimate the velocity and its ellipsoid for each random drawing with MCMC. The final derived velocities and velocity dispersions are the averaged values over the 40 random draws. Uncertainties on the velocities and their dispersions are composed of two parts: (i) the internal error from the MCMC; and (ii) the external error due to the selection of the sample. The total uncertainties of the derived values are the square root of the quadratic sum of the two parts. We compare the best fit values with the true values directly measured from the $U$, $V$, and $W$ of the individual stars (see Table \ref{tab_gcs})and find that the derived velocities and the ellipsoid are in good agreement with the true values. This confirms that the de-projection method can re-construct the 3-D kinematics for the stars within a very local volume.

\begin{table*}
\caption{Comparison of the velocity ellipsoids derived from different methods for the GCS data}\label{tab_gcs}
\begin{center}
\scriptsize
\begin{tabular}{c |c c c c c c c c c}
\hline
\hline
 &$\langle U\rangle$& $\langle V\rangle$& $\langle W\rangle$& $\langle\sigma_{U}\rangle$& $\langle\sigma_{V}\rangle$& $\langle\sigma_{W}\rangle$& $\langle\sigma_{UV}\rangle$& $\langle\sigma_{UW}\rangle$& $\langle\sigma_{VW}\rangle$\\
\hline
 direct &
-9.86$\pm$0.64&-23.49$\pm$0.46&-7.57$\pm$0.33&37.10$\pm$0.58&25.69$\pm$0.67&20.33$\pm$0.37&11.04$\pm$0.97&-6.37$\pm$1.49&4.05$\pm$2.39\\

 de-projection &-9.73$\pm$3.01&-22.84$\pm$0.74&-6.67$\pm$1.22&35.40$\pm$1.02&24.86$\pm$0.95&18.97$\pm$0.84&9.95$\pm$2.04&-8.65$\pm$5.79&3.17$\pm$6.88\\
\hline
\end{tabular}
\begin{list}{}{}
\item
The first line are the results from the direct estimation, while the second row is the results derived from the de-projection method.
\end{list}
\end{center}
\end{table*}

\subsubsection{Validation with the mock data} \label{sec_mock}
LAMOST data are neither volume complete nor do they completely cover the whole 4$\pi$ sky area. This subsequently leads to large spatial distortion in the velocity de-projection method since many lines of sight are not observed. Moreover, the LAMOST data in this work spread from 100 to 500\,pc in $|z|$, far beyond the volume of  both GCS and \emph{Hipparcos} data. The systematic bias due to the spatial variation of $\sigma$ may be stronger in this larger volume, as pointed out by \citet{MB09}.
In order to investigate both sources of systematics, we introduce a second test, assigning mock line-of-sight velocities from pre-defined velocity ellipsoids to data with the same spatial position as actual LAMOST stars.
The random draws of the line-of-sight velocities are based on a presumed Gaussian distribution function in Galactocentric cylindrical coordinates with mean velocity components of [$\langle v_R\rangle$, $\langle v_{\phi}\rangle$,  $\langle v_z\rangle$] = [0, -26, 0]\,$\kms$ and velocity dispersions of
  \begin{eqnarray} \label{eq:bolsig}
  \bolsig(R,z) = (\bolsig_{R_0} + 20z,\bolsig_{\phi_0} &+&
  20z, \bolsig_{z_0} + 30z)\kms\nonumber\\
&\times&\exp[(R_0-R)/R_\sigma],
  \end{eqnarray}
where $R_0 = 8.0 \kpc$ is the Galactocentric radius of the Sun, $R_\sigma$ is set to 5\,kpc, $\bolsig_0 \equiv  (\sigma_{R_0},\sigma_{\phi_0},\sigma_{z_0}) = (45,32,24)\kms$ following \citet{MB09}, and $z$ is in $\kpc$. The cross terms $\sigma_{R\phi}$, $\sigma_{Rz}$, and $ \sigma_{\phi z}^2$ are set to 0. For each observed star, its stellar parameters as well as the three dimensional positions are kept the same as the observed values, but its three dimensional velocity vector is randomly drawn from the Gaussian distribution function. The mock line-of-sight velocity for each star is then given by Equation~(\ref{v_los}), presuming that $U\cong-v_R$, $V\cong v_\phi$, and $W\cong v_z$ in the solar neighborhood.

\begin{figure*}
\begin{center}
\includegraphics[height=1.3\columnwidth]{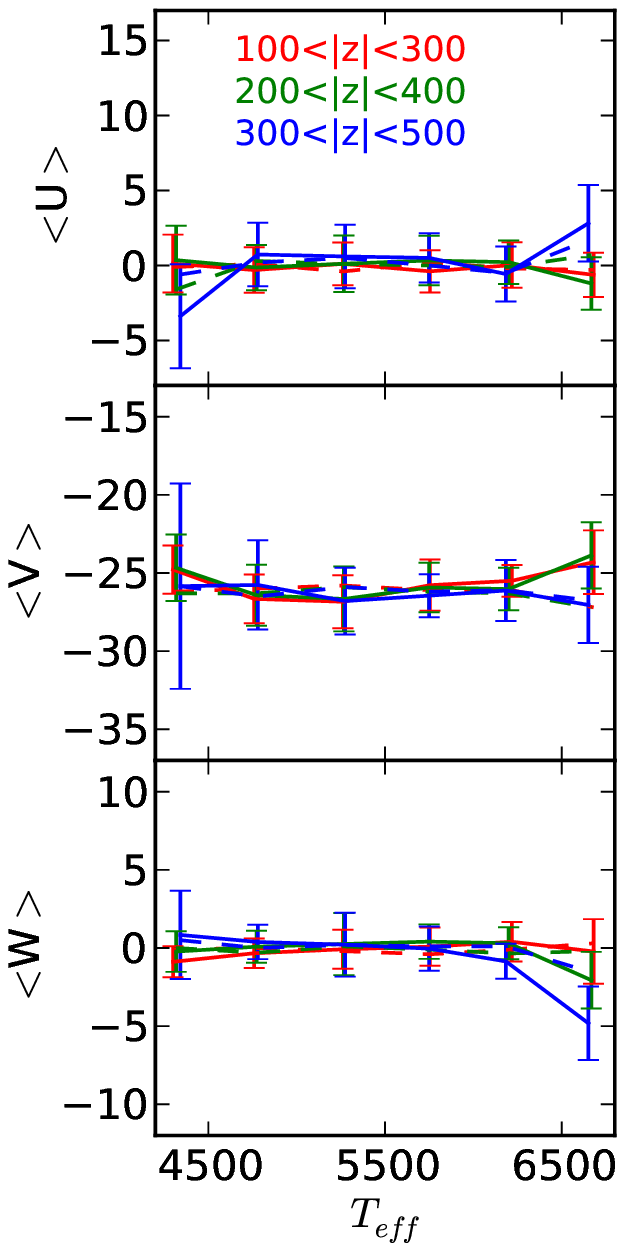}
\includegraphics[height=1.3\columnwidth]{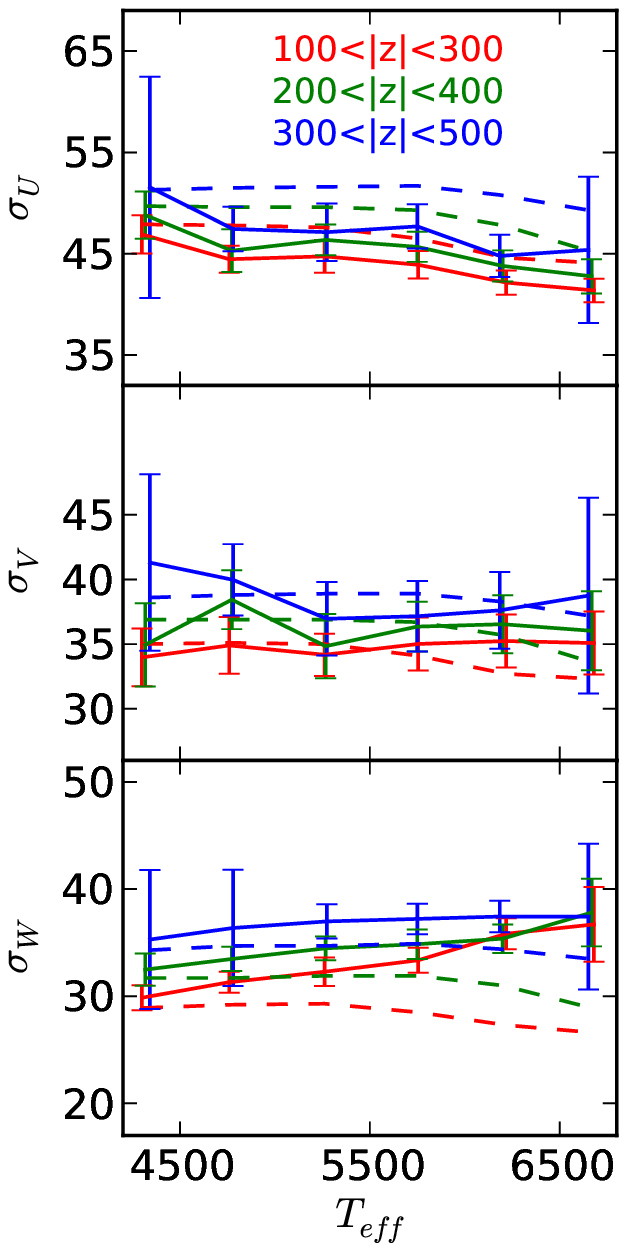}
\includegraphics[height=1.3\columnwidth]{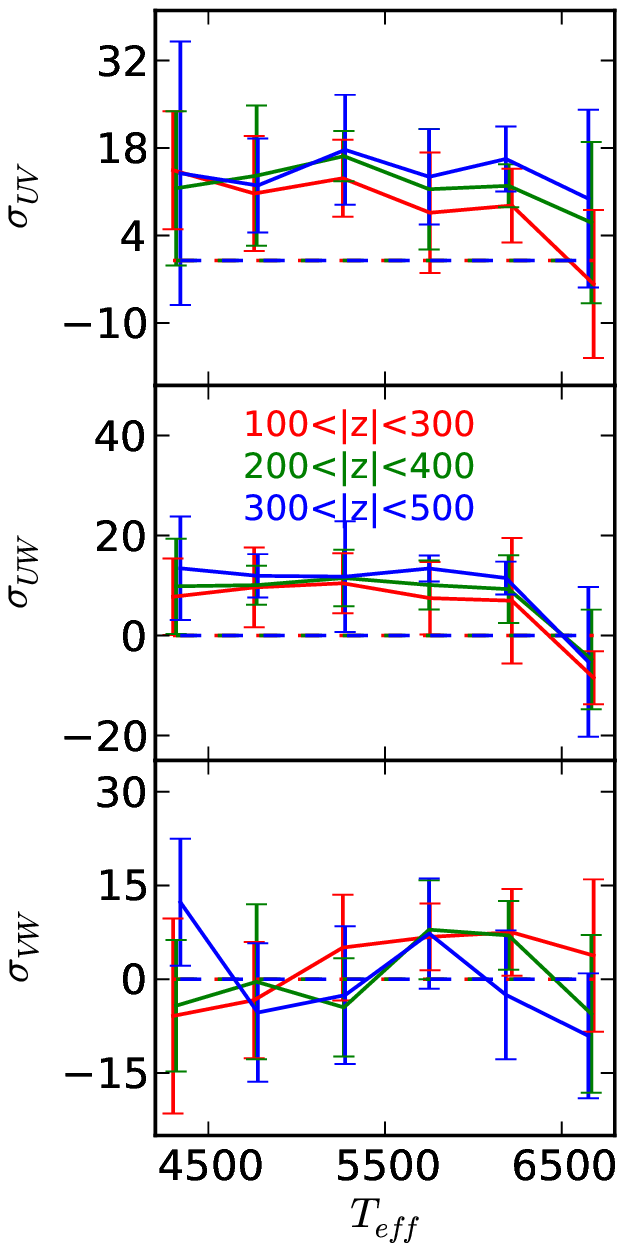}
\end{center}
\caption{Comparison of the derived velocities and velocity dispersions with their true values for one of the six mock data sets. In all panels, the solid lines represent estimated values, while the dashed lines represent the true values of the mock data. The red, green, and blue indicate samples at $100<|z|<300$, $200<|z|<400$, and $300<|z|<500$\,pc, respectively. The left panels present the three velocity components (\U, \V, and \W) in \teff\ bins in each $|z|$ slice. The second column panels show the three velocity dispersions and the right panels show the cross terms.}\label{fig_mock0}
\end{figure*}

The mock sample is first split out into three slices in height, i.e. $100\pc<|z|<300\pc$, $200\pc<|z|<400\pc$, and $300\pc<|z|<500\pc$. The overlap between neighboring slices can help to smooth the results. Then each $|z|$ slice is again split out into six bins in effective temperature with an interval of 500K ranging from 4000~K to 7000~K. Here, we assume that the mean velocity components and velocity ellipsoids are the same for stars within the same \teff\ bin in each $|z|$ slice. The results are displayed in Figure~\ref{fig_mock0} with the solid lines representing the derived values and dashed lines for the true values. Because the velocity dispersions are functions of $|z|$ and stars with different effective temperatures have different spatial distributions, the true values are not strictly flat, but show a slight bend with \teff. The reconstructed \U, \V, and \W are consistent with the true values. However, neither the velocity dispersions, nor the cross terms are in agreement with the true values. Although \citet{MB09} showed similar deviation in the cross terms, their test did not demonstrate the bias in the velocity dispersions. The inhomogeneous distribution in sky position of the LAMOST data sample (see Figure~\ref{density_distri}), which can induce even stronger geometric distortion, may be responsible for the deviation in the velocity dispersions.

\subsection{Calibration of the derived velocity ellipsoid}\label{sect:corrvel}
In general, the distortion in the velocity ellipsoid due to the unevenly distributed sample can be modeled as skew and tilt on the true velocity ellipsoid. Therefore, the distortion can be corrected by a skew and tilt model, which can be mathematically presented as matrix transformations.

The outline of the calibration is as follows. First, the mock stars are selected to be exactly located in the same spatial positions of the observed stars, combined with the simulated velocities based on a known velocity distribution function, which is only qualitatively similar to the true one.  Because the mock data are selected with the identical spatial distribution as the observed data, the true velocity ellipsoid of the mock data would be affected by the same skew and tilt as the observations. Thus, we can measure the extents of the skew and tilt from the quantitive differences between the de-projection derived velocity ellipsoids and their true values for the mock data.

In the rest of this section, we denote that all quantities with subscripts $\cdot_o$, $\cdot_m$, $\cdot_t$, and $\cdot_c$ represent for the observed values, the values measured from the mock data, the true values, and the corrected values, respectively. First, we correct the velocity vector with a simple offset:
\begin{equation} \label{offset_uvw}
\bolv_c = \bolv_o -\langle\Delta \bolv\rangle,
\end{equation}
where $\langle\Delta \bolv\rangle = \langle\bolv_m - \bolv_t\rangle$, which is derived by averaging over six simulations. Table \ref{tab_mock} lists the mean offsets at different \teff\ bins and $|z|$ slices. Most of the corrections are less than 1\,$\kms$, implying that the distortion due to the unevenly distributed sample may not severely affect the derived mean velocities from the de-projection.

Next we consider the velocity ellipsoids. For simplicity, we assume that the stars are in a three dimensional Gaussian distribution. It is worth noting that this assumption is not physically true, because the asymmetric drift of the stars in the Galactic disk skews the azimuthal velocity to the slower side \citep{BT08}. An approximation of a Gaussian distribution may derive a slower peak of the azimuthal velocity distribution. However, the derived $V$ and $\sigma_V$ based on the Gaussian approximation with the GCS data shown in Table~\ref{tab_gcs} do not show obvious bias from the true values, implying that for the real observed data in the solar neighborhood, the Gaussian approximation may not contribute significant offsets in either $V$ or $\sigma_V$. Therefore, we can still use a three dimensional Gaussian profile to approximate the velocity distribution. A detailed investigation of the effect of the Gaussian approximation in the azimuthal velocity and its dispersion is discussed in Section~\ref{sec_vphi}.

\begin{figure*}
\begin{center}
\includegraphics[height=1.3\columnwidth]{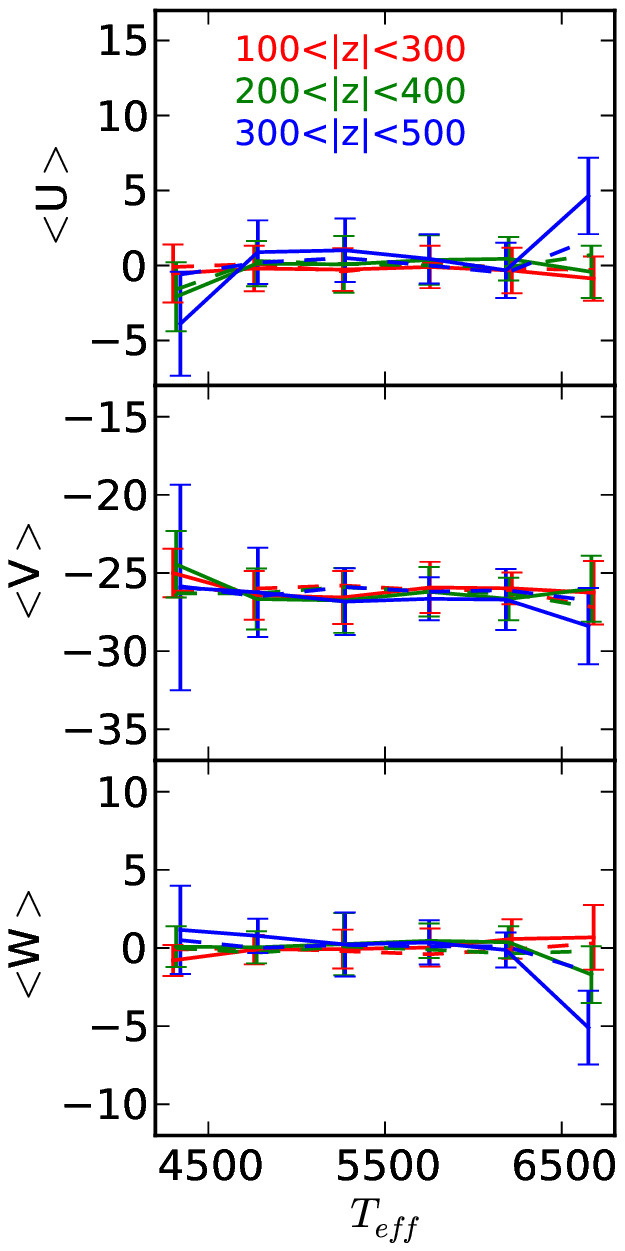}
\includegraphics[height=1.3\columnwidth]{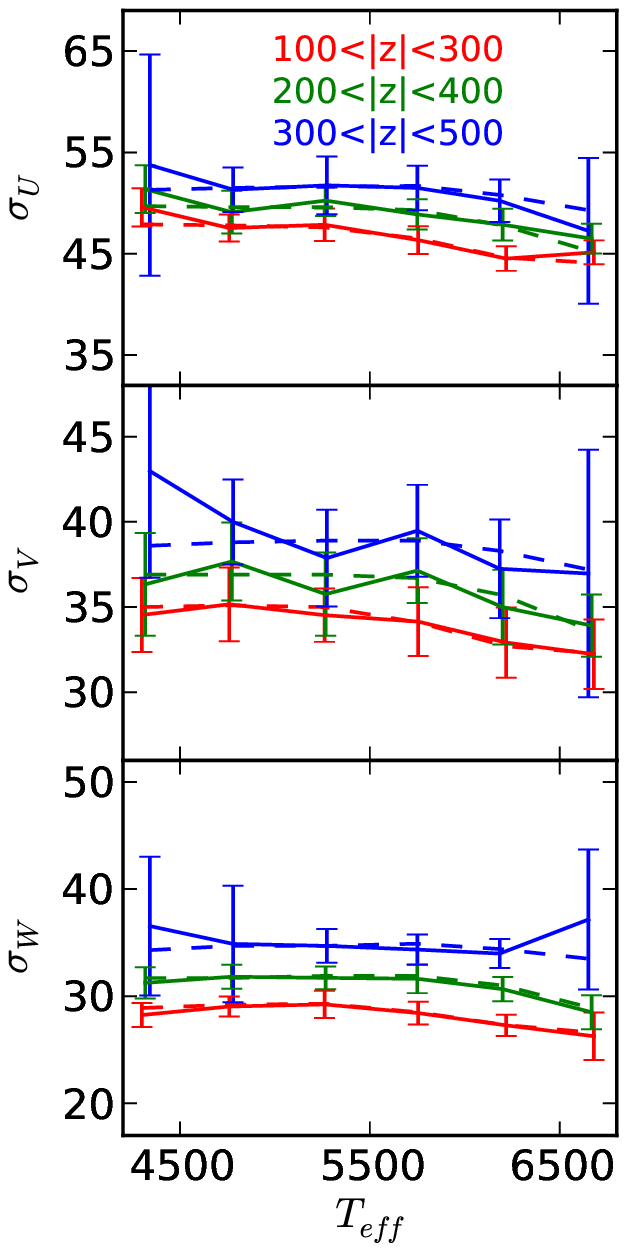}
\includegraphics[height=1.3\columnwidth]{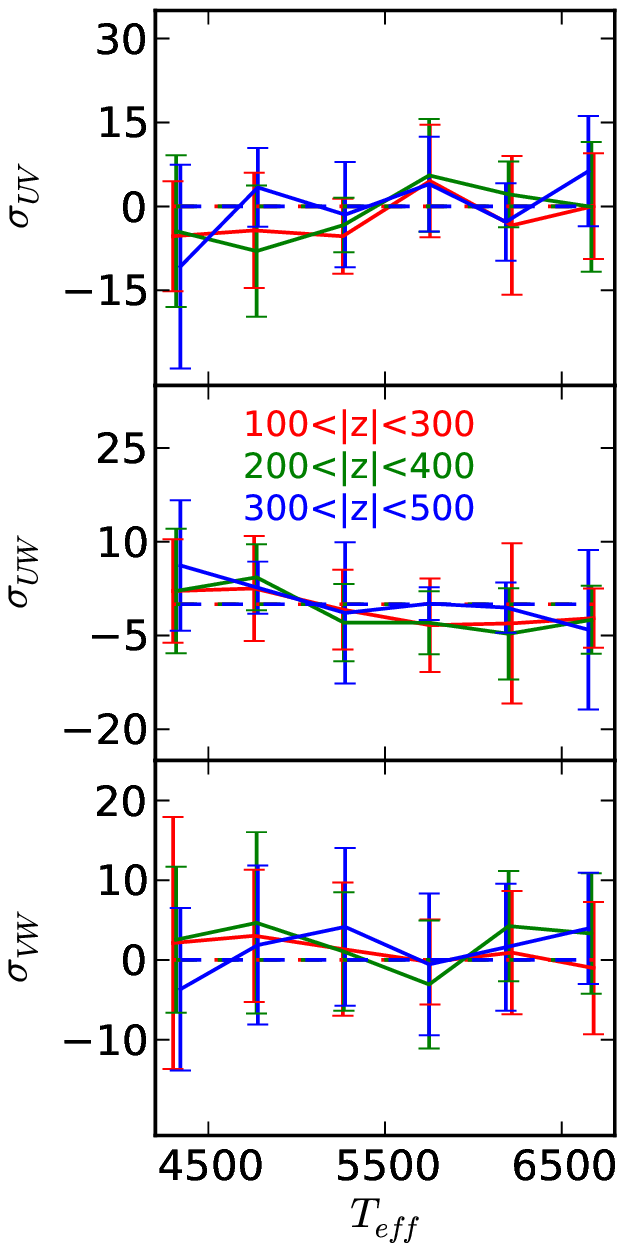}
\end{center}
\caption{Same as Figure~\ref{fig_mock0}, but the estimated values have been calibrated via the method outlined in Section~\ref{sect:corrvel}.}\label{fig_mock1}
\end{figure*}

We denote $\bolSig_t$ and $\bolSig_m$ as the true covariance matrices and the one derived with the mock data from the de-projection, respectively. Both can be rotated to align with the given axis by the following transform:
\begin{eqnarray}\label{cov_rotate}
\bolD_t&=\bolV_{t}^{-1}\bolSig_t\bolV_t\nonumber\\
\bolD_m&=\bolV_{m}^{-1}\bolSig_m\bolV_m,
\end{eqnarray}
where $\bolD_t$ and $\bolD_m$ are diagonal matrices and $\bolV_t$ and $\bolV_m$ are the rotation transform matrices. The skew due to the inhomogeneity rescales the axis ratios and hence can be quantified by the division of the two diagonal matrices.
\begin{equation}\label{scale}
\bolD_t = \bolLamb\bolD_m,
\end{equation}
where $\bolLamb$ is the scaled matrix. The tilt due to the inhomogeneity then can be quantified by the following transform:
\begin{equation}\label{tilt}
\bolSig_t=\bolV_t\bolLamb\bolV_m^{-1}\bolSig_m\bolV_m\bolV_t^{-1}.
\end{equation}
We define the correction matrices as $\bolA=\bolV_t\bolLamb\bolV_m^{-1}$ and $\bolB=\bolV_m\bolV_t^{-1}$. Then, the observed velocity covariance $\bolSig_o$ can be calibrated by
\begin{equation}\label{corr}
\bolSig_c = \bolA\bolSig_o\bolB.
\end{equation}
The calibration matrices can be derived from the simulations with mock data combining the observed 3-D positions with arbitrarily drawn velocities from given velocity ellipsoids. In practice, the random mock data may introduce some statistical fluctuations in $\bolA$ and $\bolB$. Therefore, we run six simulations to generate six sets of $\bolA$ and $\bolB$. The final calibrated velocity ellipsoid is averaged over the calibration by the six sets of transform matrices. Figure~\ref{fig_mock1} shows the calibrated results, which are now consistent with their true values after the correction. It shows that not only the dispersions but also most of the cross terms have been well reconstructed by the calibration. The calibration corrects the systematics raised from both the unevenly spatial distribution of the observed sample and the spatial variation of $\sigma$.

\begin{table}
\caption{The mean offsets of velocity to calibrate the geometric distortion}\label{tab_mock}
\begin{center}
\scriptsize
\begin{tabular}{c  c c c  c }
\hline
\hline
\hline
Teff&$\langle N\rangle$& $\langle\Delta U\rangle$& $\langle\Delta V\rangle$& $\langle\Delta W\rangle$\\
\hline
\hline
&&\multicolumn{3}{c}{100$<|z|<$300 pc}\\
\hline
4250&13229&0.66$\pm$0.68&0.22$\pm$0.89&-0.09$\pm$0.41\\
4750&19593&-0.10$\pm$0.25&-0.23$\pm$0.29&-0.23$\pm$0.34\\
5250&25805&0.38$\pm$0.38&-0.27$\pm$0.40&-0.01$\pm$0.32\\
5750&33453&-0.29$\pm$0.20&0.15$\pm$0.42&0.06$\pm$0.37\\
6250&23565&0.36$\pm$0.31&0.48$\pm$0.30&-0.18$\pm$0.53\\
6750&6739&0.24$\pm$0.79&1.96$\pm$1.29&-0.90$\pm$1.44\\
\hline
\hline
&&\multicolumn{3}{c}{200$<|z|<$400 pc}\\
\hline
4250&7600&2.44$\pm$0.92&-0.22$\pm$1.26&-0.31$\pm$0.25\\
4750&16639&-0.26$\pm$0.28&0.23$\pm$0.73&0.05$\pm$0.40\\
5250&27523&0.01$\pm$0.32&0.10$\pm$0.52&0.01$\pm$0.23\\
5750&36303&-0.03$\pm$0.27&0.27$\pm$0.34&-0.05$\pm$0.33\\
6250&18621&-0.22$\pm$0.48&0.63$\pm$0.63&-0.06$\pm$0.46\\
6750&3436&-0.77$\pm$0.92&2.13$\pm$1.29&-0.15$\pm$1.42\\
\hline
\hline
&&\multicolumn{3}{c}{300$<|z|<$500 pc}\\
\hline
4250&3797&0.58$\pm$2.11&0.04$\pm$1.17&-0.30$\pm$0.48\\
4750&11443&-0.16$\pm$0.43&0.47$\pm$0.76&-0.38$\pm$0.51\\
5250&22503&-0.41$\pm$0.50&0.04$\pm$0.59&-0.01$\pm$0.26\\
5750&33595&0.06$\pm$0.27&0.20$\pm$0.40&-0.40$\pm$0.30\\
6250&13986&-0.24$\pm$0.91&0.58$\pm$0.44&-0.73$\pm$0.29\\
6750&1566&-1.81$\pm$2.24&1.36$\pm$1.83&0.44$\pm$2.09\\
\hline
 \hline
\hline
\end{tabular}
\begin{list}{}{}
\item
The $\langle\Delta U\rangle$, $\langle\Delta V\rangle$ and $\langle\Delta W\rangle$ are the average offsets of the velocities for various \teff\ bins at different $|z|$ slice.
\end{list}
\end{center}
\end{table}

\section{The results of the LAMOST DR1 data}\label{sec_result}
\begin{figure*}
\begin{center}
\includegraphics[height=1.3\columnwidth]{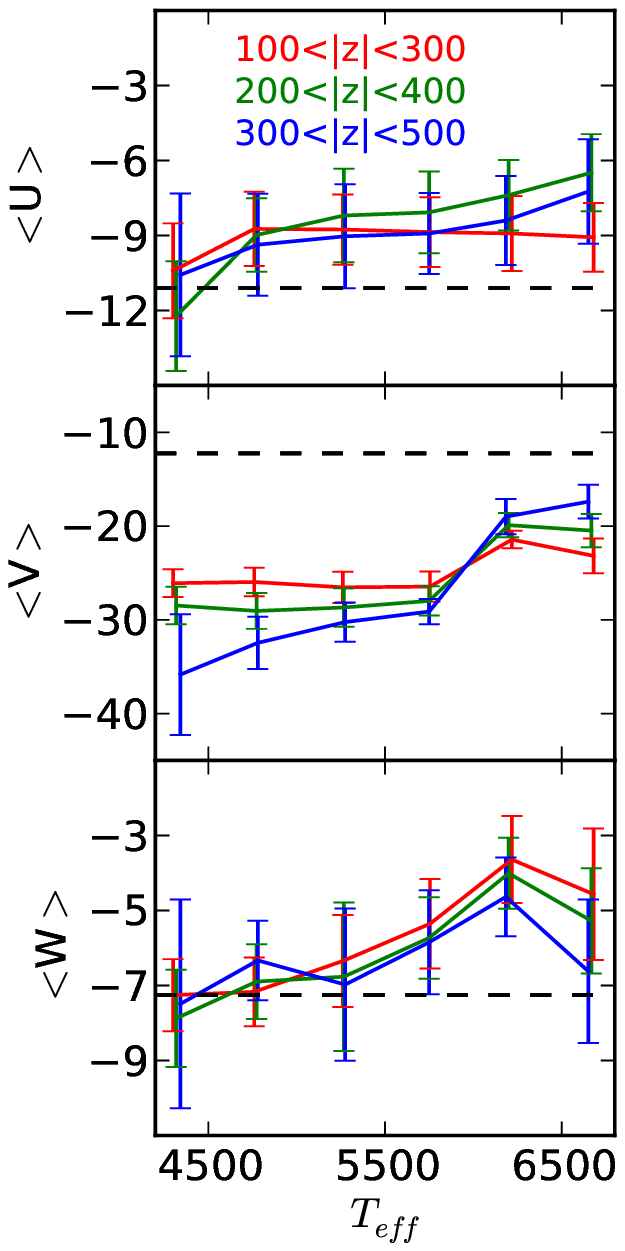}
\includegraphics[height=1.3\columnwidth]{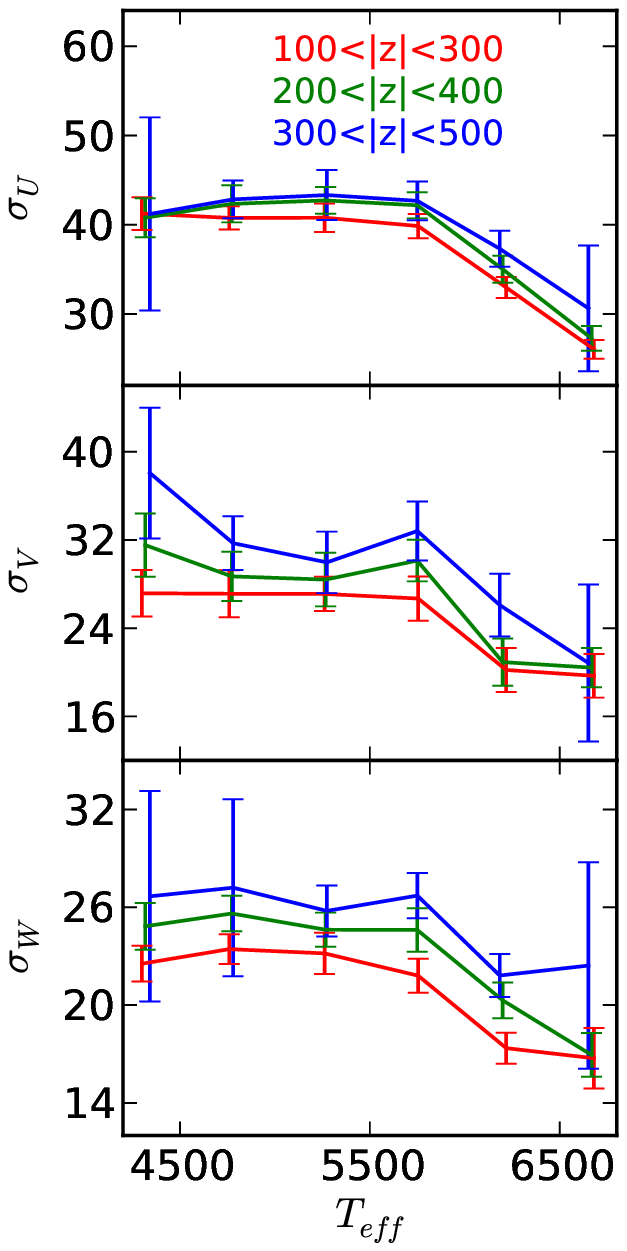}
\includegraphics[height=1.3\columnwidth]{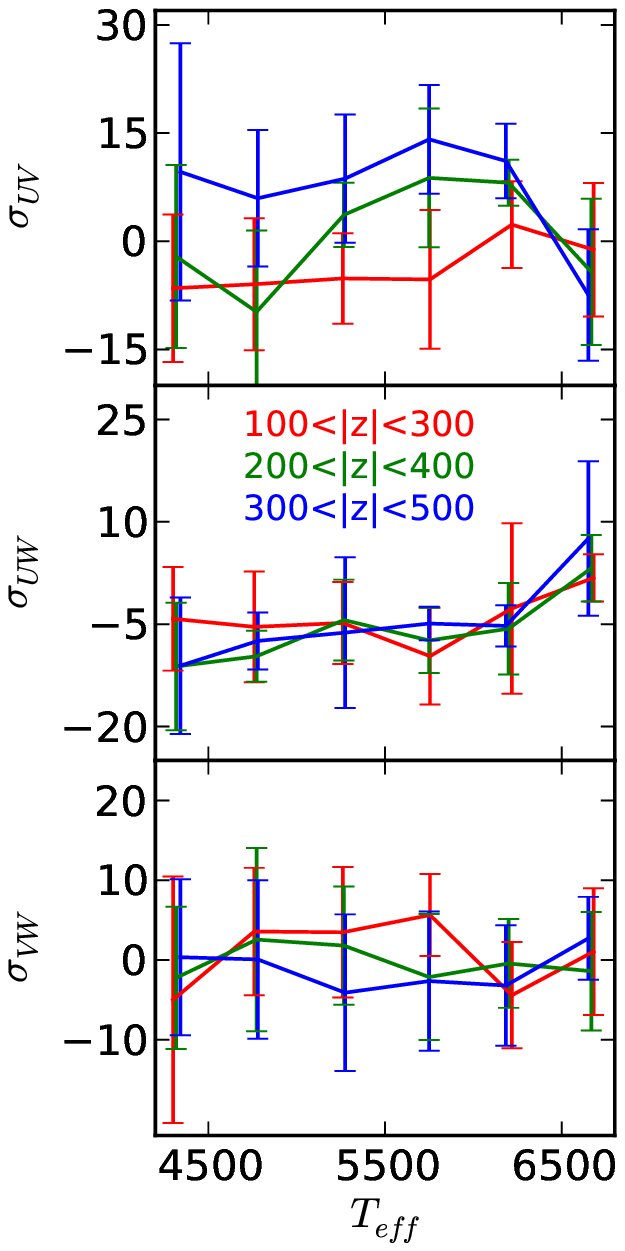}
\end{center}
\caption{The mean velocities and velocity ellipsoids estimated from the LAMOST DR1 samples. The symbols are same as Figure~\ref{fig_mock0}. The three black dash lines in the left panels indicate the zero points of the three components with respect to the LSR according to \citet{SBD10}.}\label{fig_UVW_Teff_az}
\end{figure*}

We apply the calibrated de-projection approach to the LAMOST data. Figure~\ref{fig_UVW_Teff_az} and Table~\ref{tab_sigma2b} show the results. The correlations of the three velocity and the six velocity ellipsoid components with \teff\ are demonstrated in three slices of $|z|$, i.e. $100\pc<|z|<300\pc$ (red), $200\pc<|z|<400\pc$ (green), and $300\pc<|z|<500\pc$ (blue). The data at each $|z|$ slice are split out into 6 \teff\ bins between 4000$<$\teff$<$7000\,K with an interval of 500\,K. Similar to the procedure for GCS data described in section~\ref{sect:gcsvalid}, the uncertainty of the velocity and its ellipsoid is the square root of the quadratic sum of two parts: (i) the internal error from MCMC; and (ii) the external error from the selection of the sample data on the sky. To measure the error from the second part, we run a 40-time bootstrap to randomly resample the observed data on the sky. For the mean velocities, the uncertainty is mainly from the first part, while for the second moments, the uncertainties are dominated by the second effect. In this sense, the final uncertainty of the velocity ellipsoid cannot be effectively reduced until the LAMOST survey has covered a large, contiguous region of the northern sky, as it will by the end of the mission.

\subsection{The mean velocities}\label{sec_uvw}
The left panels of Figure \ref{fig_UVW_Teff_az} present the results of the three components of the average velocity (\U, \V, and \W).

The most obvious feature is that the three mean velocities shown in the left panels are all correlated with effective temperature for all $|z|$ bins.  For the radial velocity \U\ in the top-left panel, at \teff=4250\,K, \U\ is around $10-12$\,$\kms$, which is at around 0\,$\kms$ with respect to the local standard of rest (LSR, the black dashed line in the panel) adopting the solar motion of $U_{\odot}=11.1$\,$\kms$ \citep{SBD10}. In other \teff\ bins, \U\ for the stars within $100<|z|<300$\,pc are approximately flat at around $-9.07$ to $-8.74$\,$\kms$. For the stars in $200<|z|<400$ ($300<|z|<500$), on the other hand, \U\ increases from $-8.98\pm1.47$ ($-9.37\pm2.04$)\,$\kms$ at \teff$=4750$\,K to $-6.49\pm1.54$ ($-7.24\pm2.09$)\,$\kms$ at \teff$=6750$\,K. Although the change of \U\ from cool to warm is small, the increasing trend is quite clear, especially when \teff$>6000$\,K. The direction of \U\ is toward the Galactic center; thus the top-left panel of Figure~\ref{fig_UVW_Teff_az} indicates that either the warmer stars tend to move inward the Galaxy or the cooler stars tend to move outward.

For the azimuthal velocity \V\ in the middle-left panel, the stars show a clear break at \teff$=6000$\,K. The stars cooler than 6000\,K are located at around -30\,$\kms$, while those warmer than 6000\,K go abruptly up to around -20\,$\kms$. This is another version of Parenago's discontinuity \citep{parenago1950} and has been re-discovered by DB98 at $B-V=0.61$, which corresponds to \teff$\sim6000$\,K according to \citet[][]{SF00}. The cooler stars are on average older than the warmer stars and hence most of them have had a longer time to experience scattering, which can increase the velocity dispersion and therefore increase the asymmetric drift \citep{Jenkins1992, BT08}. This leads to a slower azimuthal velocity for the older (cooler) stars. Another interesting feature is that the LAMOST data shows an unusual gradient along $|z|$ for warmer stars. For the stars cooler than 6000\,K, the azimuthal velocity decreases with $|z|$, while it turns out to increase with $|z|$ for the stars warmer than 6000\,K. The former trend is natural since the stars at higher $|z|$ need more vertical energy and hence slightly lose their angular momenta and rotate slower, given that the stars are in equilibrium. However, it is difficult to understand why the warmer, or younger, stars show the trend turning around.

As did the other two velocity components, the vertical velocity \W\ also shows a correlation with effective temperature. Compared to the zero velocity with respect to the LSR (shown as the black dashed horizontal line at 7.25\,$\kms$ \citep{SBD10}), the stars with \teff$<5500$\,K roughly stay around the zero point. However, the warmer stars with \teff$>5500$\,K show higher \W\ than the zero point, i.e., they are moving toward the Galactic north pole up to \W$\sim-4$\,$\kms$, which is equivalent to +3\,$\kms$ with respect to LSR. The stars located at larger $|z|$ seem to have a smaller \W\ than those located in lower $|z|$, although the trend is not quite statistically significant.

\begin{table}
\caption{The solar motion with respect to the LSR.}\label{tab:lsr}
\centering
\begin{tabular}{c|c}
\hline\hline
& $\kms$\\
\hline
$U_\odot$ 
& 9.58$\pm$2.39\\
\multirow{2}{*}{$V_\odot$}
& 10.52$\pm$1.96 ($h_R$=2.6\,kpc, L=5.2\,kpc)\\
& 10.05$\pm$1.98 ($h_R$=2.5\,kpc, L=5.0\,kpc)\\
& 13.09$\pm$1.85 ($h_R$=3.0\,kpc, L=6.0\,kpc)\\
$W_{\odot}$ & 7.01$\pm$1.67\\
\hline\hline
\end{tabular}
\end{table}

The stars cooler than 6000\,K over all $|z|$ bins are on average old. Since they do not show a significant gradient in \V\, it is safe to assume that they are in equilibrium. Then, they can be used to determine the solar motion with respect to LSR. The solar motion in radial and vertical directions can be derived from
\begin{eqnarray}
U_{LSR}&=-\langle U\rangle\nonumber\\
W_{LSR}&=-\langle W\rangle.
\end{eqnarray}
Thus, we directly obtain $U_{\odot}=9.58\pm2.39$\,$\kms$ and $W_{\odot}=7.01\pm1.67$\,$\kms$ by averaging over the \U\ and \W\ listed in the 3rd and 5th columns of Table~\ref{tab_sigma2b} for stars below 6000\,K in all $|z|$ slices.

In order to estimate $V_{\odot}$, the asymmetric drift is required and a dynamical model has to be introduced. We start from Equation (4.228) of \citet{BT08}, which gives the approximation of the asymmetric drift from the circular speed, velocity dispersions, and the stellar density.
\begin{equation}\label{eq:strom}
  \overline{\upsilon}_s-\Vsun =\va\simeq
  \frac{\overline{\upsilon^2_{\!R}}}{2\upsilon_{\mathrm{c}}}
  \left[\frac{\sigma^2_{\phi}}{\vrq} - 1 -
    \frac{\partial\ln(\nu\vrq)}{\partial\ln R} -
    \frac{R}{\vrq} \frac{\partial(\overline{\upsilon_R\upsilon_z})} {\partial z}
  \right],
\end{equation}   
where $R$ is Galactocentric radius, $z$ the height above the
plane, $\upsilon_{\mathrm{c}}$ the circular speed, and $\nu$ the number
density of stars, while a bar indicates the average value. Because the data are very close to the Galactic mid-plane, the cross term of $\overline{v_Rv_z}$ does not significantly vary with $|z|$ \citep{BvW14} and therefore the last term of the equation is very small and can be negligible. Assume that the stellar density $\nu$ and $\sigma_U$ are exponential functions of $R$:
\begin{eqnarray}
\nu&\sim exp(-{R\over{h_R}})\\
\sigma_U^2&\sim exp(-{R\over{L}}),
\end{eqnarray}
where $h_R$ and $L$ are scale length for the stellar density and $\sigma_U$, respectively. Then Equation (4.228) can be rewritten as:
\begin{equation}\label{eq:va}
v_a\cong{1\over{2v_c}}[\sigma_V^2+({1\over{h_R}}+{1\over{L}}-{1\over{R}})R\sigma_U^2].
\end{equation}
Therefore, the solar motion in $V$ can be estimated from
\begin{equation}\label{eq:vlsr}
V_{\odot}=-(\langle V\rangle+v_a).
\end{equation}

Here, we set $h_R=2.6$\,kpc following \citet{Juricetal2008_short}  or $h_R=2.5$\,kpc following \citet{schoenrich12} and \citet{Sharma14}. According to \citet{kf11}, $L$ is equal to $2h_R$ given an isothermal disk. Although this relation has not been confirmed for the case of our Galaxy, it is a reasonable guess. As comparison, the different case of $h_R=3.0$\,kpc and $L=6.0$\,kpc is also considered. The final results, including all three components, are listed in Table~\ref{tab:lsr}. Our estimates of the solar motion in three dimensions are in agreement with \citet{SBD10} and~\citet{huang15}.


\begin{table}[htbp]
\caption{The vertical gradient of the velocity dispersions at each \teff\ bin.}\label{tab:vertgrad}
\centering
\begin{tabular}{c|ccc}
\hline
\hline
\teff\ & $d\sigma_U/d|z|$ &$d\sigma_V/d|z|$ &$d\sigma_W/d|z|$ \\
K & $\kms{\rm kpc}^{-1}$ & $\kms{\rm kpc}^{-1}$ & $\kms{\rm kpc}^{-1}$ \\
\hline
4250 & -0.2$\pm$1.4 & 54.5$\pm$3.4 & 20.7$\pm$0.7 \\
4750 & 10.4$\pm$1.7 & 22.9$\pm$2.3 & 18.9$\pm$1.0 \\
5250 & 12.7$\pm$2.1 & 14.3$\pm$0.4 & 13.1$\pm$0.5 \\
5750 & 14.1$\pm$2.9 & 30.7$\pm$1.2 & 24.6$\pm$1.1 \\
6250 & 21.8$\pm$0.4 & 29.4$\pm$7.0 & 22.3$\pm$2.2 \\
6750 & 23.0$\pm$3.4 & 5.8$\pm$0.5 & 28.4$\pm$8.3 \\
\hline
\hline
\end{tabular}
\end{table}

\subsection{The velocity ellipsoids}\label{sec_uvw}
The second column panels of Figure \ref{fig_UVW_Teff_az} show the velocity dispersions as functions of \teff\ and $|z|$, and the right panels show the derived cross terms.
The velocity dispersions show clear trends either along \teff\ or $|z|$. First, all three velocity dispersions show an abrupt drop at around \teff$\sim6000$\,K. This sudden drop has been seen in all $|z|$ bins. It is in agreement with the turn-around point in \V\ shown in the middle-left panel in Figure~\ref{fig_UVW_Teff_az}. The decreased velocity dispersions for stars with \teff$>6000$\,K again imply that the young stars are less affected by scattering and hence are kinematically cooler.

Second, the velocity dispersions show vertical gradients in most of the \teff\ bins. Particularly, the vertical gradient of $\sigma_U$ is correlated with \teff, as shown in the second column of Table~\ref{tab:vertgrad}. Although the vertical gradients of $\sigma_V$ and $\sigma_W$ are mostly larger that those of $\sigma_U$, they do not show clear trends along \teff\,(also see Table~\ref{tab:vertgrad}).

\begin{figure}[htbp]
\centering
\includegraphics[height=1.0\columnwidth]{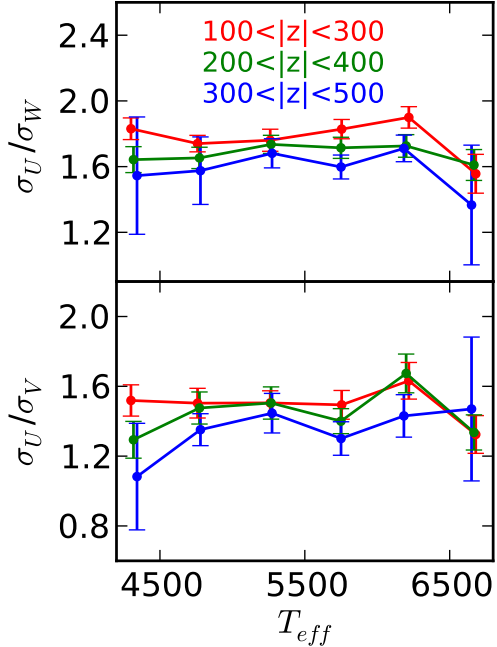}
\caption{ The axis ratios of the ellipsoids in various temperature bins at each $|z|$ slice.}\label{fig_ratio2}
\end{figure}

The ratios of the velocity dispersions shown in Figure~\ref{fig_ratio2} indicate the shape of the velocity ellipsoids. Columns 12 and 13 in Table~\ref{tab_sigma2b} also list the values of the two ratios. Both the ratios drop with $|z|$ in most \teff\ bins. The only exception is at \teff$=6750$\,K, in which the ratios seem not correlated with $|z|$ (Figure~\ref{fig_ratio2}).

\begin{figure}[htbp]
\begin{center}
\includegraphics[height=1.0\columnwidth]{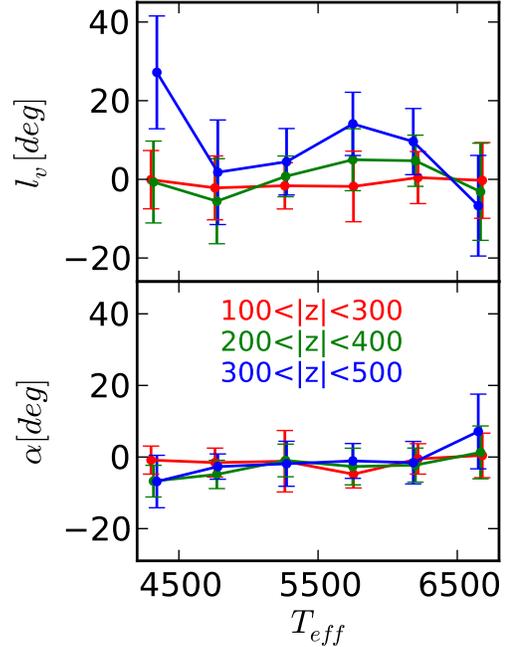}
\end{center}
\caption{The vertex deviation $l_v$ and tilt angle $\alpha$ in various effective temperature bins at each $|z|$ slice.}\label{fig_lv_az}
\end{figure}

With derived cross terms, we can also estimate the vertex deviation and the tilt angle at each \teff\ bin and $|z|$ slice. The definition of the vertex deviation is
\begin{equation} \label{lv}
{l_v} \equiv {1\over2}{\rm arctan}({2\sigma_{UV}^2\over{\sigma_{U}^2 - \sigma_{V}^2}}),
\end{equation}
and the definition of the tilt angle is
\begin{equation} \label{alpha}
{\alpha} \equiv {1\over2}{\rm arctan}({2\sigma_{UW}^2\over{\sigma_{U}^2 - \sigma_{W}^2}}).
\end{equation}
The right panels of Figure~\ref{fig_UVW_Teff_az} show the derived cross terms of the velocity ellipsoids. Their values are listed in Table~\ref{tab_sigma2b}. Figure~\ref{fig_lv_az} shows the estimated $l_v$ and $\alpha$ as functions of \teff\ and $|z|$.

We find that $l_v$ varies with $|z|$. It is around  0$^\circ$ for stars with \teff$<6500$\,K at $100<|z|<300$ and $200<|z|<400$\,pc and then it becomes positive at $300<|z|<500$\,pc. The vertex deviation for stars with \teff$>6500$\,K, as an exception, changes with $|z|$ in the opposite sense, i.e., it decreases with $|z|$ from zero to $-6.7^\circ$, although the uncertainty is very large.
On the other hand, the tilt angle does not show clear trends with either \teff\ or $|z|$, which is different with \citet[][hereafter B14]{B14}, who found the tilt angle changes for the hot dwarfs compared to the giants or cool dwarfs using RAVE~\citep{rave} data. The reason of the discrepancy is hard to be investigated, because of the large difference of the volumes between the LAMOST and RAVE data and the difference in methodologies. 

\begin{table*}
\caption{Comparison of the velocity dispersions between DB98 and this work.}\label{tab:compareDB98}
\centering
\begin{tabular}{c|c|cccc|cccc}
\hline\hline
\teff\ & $B-V$ & \multicolumn{4}{|c|}{DB98} & \multicolumn{4}{|c}{This work ($100<|z|<300$)}\\
\hline
&&$\sigma_1$&$\sigma_1/\sigma_2$&$\sigma_1/\sigma_3$&$l_v$&$\sigma_1$&$\sigma_1/\sigma_2$&$\sigma_1/\sigma_3$&$l_v$\\
\hline
K & mag & \multicolumn{3}{|l}{$\kms$} & degree & \multicolumn{3}{|l}{$\kms$} & degree\\
\hline
5250 & 0.719 & 37.20$^{+1.41}_{-0.93}$ & 1.44$^{+0.12}_{-0.12}$ & 2.04$^{+0.61}_{-0.16}$ & 13.1$^{+6.7}_{-7.6}$ & 40.81$\pm$5.06 &1.51$\pm$0.29 & 1.76$\pm$2.05  & -1.63$\pm$5.90\\
5750 & 0.582 & 37.64$^{+1.37}_{-0.94}$ & 1.61$^{+0.07}_{-0.18}$ & 1.78$^{+0.48}_{-0.04}$ & 10.2$^{+5.6}_{-6.0}$ & 39.97$\pm$5.11 &1.50$\pm$0.33 & 1.86$\pm$1.21 & -1.80$\pm$8.98\\
6250 & 0.525 & 32.93$^{+1.09}_{-0.75}$ & 1.51$^{+0.13}_{-0.12}$ & 2.19$^{+0.64}_{-0.19}$ & 1.9$^{+6.0}_{-6.1}$ & 32.97$\pm$5.77 &1.62$\pm$0.40 & 1.91$\pm$3.11 & 0.45$\pm$6.62\\
6750 & 0.412 & 26.26$^{+0.80}_{-0.59}$ & 1.66$^{+0.12}_{-0.15}$ & 2.16$^{+0.52}_{-0.15}$ & 10.2$^{+5.0}_{-5.4}$& 26.05$\pm$5.47 &1.32$\pm$0.44 & 1.56$\pm$2.92 & -0.27$\pm$9.63\\
\hline
\hline
\end{tabular}
\end{table*}

\begin{table*}
\caption{Comparison of the velocity dispersions between B14 and this work.}\label{tab:compareB14}
\centering
\begin{tabular}{c|c|ccc|ccc}
\hline\hline
Height\ & \teff & \multicolumn{3}{|c|}{B14} & \multicolumn{3}{|c}{This work}\\
\hline
&&$\sigma_1$&$\sigma_1/\sigma_2$&$\sigma_1/\sigma_3$&$\sigma_1$&$\sigma_1/\sigma_2$&$\sigma_1/\sigma_3$\\
\hline
pc & K & \multicolumn{3}{|c}{$\kms$} & \multicolumn{3}{|c}{$\kms$}\\
\hline
\multirow{2}{*}{$100<|z|<300$} 
& $>$6000 & 30.47$\pm$0.70& 1.52$\pm$0.11 & 1.95$\pm$0.08 & 29.51$\pm$4.89
 &1.47$\pm$0.21 & 1.73$\pm$0.25\\
& $<$6000&38.42$\pm$2.30 & 1.52$\pm$0.05 & 1.71$\pm$0.05 & 40.71$\pm$0.54 &1.51$\pm$0.01 & 1.80$\pm$0.05\\
\hline
\multirow{2}{*}{$200<|z|<400$} 
& $>$6000 & 32.70$\pm$1.54 & 1.46$\pm$0.19 &1.80$\pm$0.08& 31.20$\pm$5.53 &1.51$\pm$0.25 & 1.67$\pm$0.09\\
& $<$6000& 43.49$\pm$2.77 & 1.57$\pm$0.04 & 1.75$\pm$0.01 & 42.14$\pm$0.80 & 1.43$\pm$0.09 & 1.70$\pm$0.03\\
\hline
\multirow{2}{*}{$300<|z|<500$} 
& $>$6000 &36.57$\pm$0.70 & 1.40$\pm$0.18 & 1.69$\pm$0.03 & 34.23$\pm$4.78 &1.42$\pm$0.06 &1.61$\pm$0.16\\
& $<$6000&48.76$\pm$2.50 & 1.63$\pm$0.04 & 1.67$\pm$0.07 & 42.83$\pm$0.76 &1.32$\pm$0.15 & 1.62$\pm$0.05\\
\hline
\hline
\end{tabular}
\end{table*}

\subsection{Comparison of the velocity ellipsoids with other works}\label{compare}
Although the volume of our sample at $100<|z|<300$\,pc does not exactly overlap with that of the sample of DB98, which only extends to 70\,pc, we will compare our sample to that of DB98 under the assumption that the velocity dispersions do not largely change from $|z|<$100 to 300\,pc. The color index $B-V$ used by DB98 can be converted into effective temperature according to \citet{SF00}. Specifically, the data at around $B-V=0.412$ (the color index bin 4 in DB98) corresponds to \teff$\sim6600$\,K, the data at around $B-V=0.525$ (bin 6 in DB98) corresponds to \teff$\sim6200$\,K, the data at around $B-V=0.582$ (bin 7 in DB98) corresponds to \teff$\sim5800$\,K, and the data at $B-V=0.719$ (bin 9 in DB98) corresponds to \teff$\sim5200$\,K. Because DB98 did not directly show the dispersions along the $U$, $V$, and $W$ directions, but showed the dispersions along the major axis of the velocity ellipsoids and the ratio of the major axis to the middle and minor axis, we also rotate the velocity ellipsoid showing in Table~\ref{tab_sigma2b} and find the longest, middle, and shortest axes of the derived ellipsoids to compare with DB98. Table~\ref{tab:compareDB98} shows the comparison of the largest velocity dispersion $\sigma_1$ and the two ratios of $\sigma_1/\sigma_2$ and $\sigma_1/\sigma_3$ between DB98 and this work. In most cases, the velocity dispersions with roughly the same \teff\ are slightly larger in our data than those in DB98 by up to $\sim3$\,$\kms$. This is reasonable since the height of the data in this work is slightly larger than those in DB98.

It also shows that the ratio $\sigma_1/\sigma_2$ derived in this work is on average consistent with DB98, while $\sigma_1/\sigma_3$ is slightly smaller than DB98. In other words, compared with DB98, the velocity ellipsoid in this work is slightly broader in the $z$ direction, which is probably because the LAMOST data extend over a larger range in $|z|$. Notice that the errors in axis ratios are larger in this work. These errors may be overestimated because they are propagated from the large errors in the cross terms via the rotation transforms, rather than deriving directly from the observed data.

Surprisingly, for the stars located at $100<|z|<300$\,pc, the vertex deviation is around zero for all \teff\ bins. This is inconsistent with the results in DB98. First, it is noted that Sections~\ref{sect:gcsvalid} and~\ref{sec_mock} have shown that the cross terms of the velocity ellipsoids are the easiest affected in the velocity de-projection method, as \citet{MB09} have stated. Both the method in DB98 and in this work are based on the velocity de-projection. Therefore, both the results of the vertex deviation are possibly affected by systematic bias. Because DB98 de-projected the velocity components from the tangental velocities measured from proper motions, while we derive them from the line-of-sight velocities, the systematic bias in both works may go toward different directions. Second, even though we have calibrated the velocity ellipsoid, the corrected $\sigma_{UW}$ may still slightly bias from their true values (see the top-right panels of Figure~\ref{fig_mock1}). As a consequence, it is quite hard to compare the vertex deviation between this work and DB98. This issue cannot be easily solved until the 3 dimensional velocities of stars are provided.

B14 analysed the kinematics of $\sim$400 000 RAVE stars, and decomposed the sample into hot and cold dwarfs, red-clump and non-clump giants. For each of these classes, B14 provided an analytic model for the velocity ellipsoid at each point in the (R, z) plane. With the help of this model, we compared the velocity dispersions between B14 and this work. Table~\ref{tab:compareB14} shows the comparison of the largest velocity dispersion $\sigma_1$ and the two ratios of $\sigma_1/\sigma_2$ and $\sigma_1/\sigma_3$ between B14 and this work in the different layers. The $\sigma_1$ , $\sigma_1$/$\sigma_2$, and $\sigma_1$/$\sigma_3$ between the two works are in good agreement with each other within 1$\sigma$ for most cases. However, for stars with \teff$>6000$\,K and $200<|z|<400$ and those with \teff$<6000$ and $300<|z|<500$, the two works have differences by about 2-$\sigma$.

It is not easy to compare the dispersions with S12 since they separate the data into metallicity bins rather than temperature or color index. Moreover, most of the data in S12 are at heights larger than 500\,pc, which does not overlap with our data. However, a qualitative comparison is still worthwhile since S12 directly used three velocity components to derive the velocity ellipsoids. First, we compare the velocity dispersions at $300<|z|<500$\,pc and \teff$=6250$\,K (see Table~\ref{tab_sigma2b}) with the data at $-0.5 <$~[Fe/H]~$< 0.2$ and $\langle z\rangle=-0.69$\,kpc in S12, which have $\sigma_R=38.8^{+1.2}_{-1.3}$\,$\kms$, $\sigma_\phi=27.7^{+0.4}_{-0.4}$\,$\kms$, and $\sigma_z=22.4^{+0.7}_{-0.7}$\,$\kms$, and find that they agree with each other. Then, we compare $l_v=9.60^\circ\pm2.56^\circ$ with their result, which is $8.3^\circ\pm3.4^\circ$. These two quantities are again consistent with each other.  We also compare the tilt angle $\alpha=-1.73^\circ\pm0.76^\circ$ with S12, which obtained $-5.6^\circ\pm2.1^\circ$. Although the direction of the tilt angles are same, the values have about $2\sigma$ difference. This is probably because the tilt angle increases very rapidly with $|z|$ \citep{BvW14} and the data in S12 are located in larger $|z|$ than those in this work.

\section{Discussion}\label{sect:discussion}
\subsection{The effect of the non-Gaussianity of $V$}\label{sec_vphi}
It is well known that the $V$ distribution is not Gaussian but has a long tail to the slow side; hence, the Gaussian approximation in Equation~(\ref{likelihood}) may not be correct. It is worthwhile to investigate the systematic bias in the derived velocity ellipsoids due to the Gaussian approximation.

\begin{table}
\caption{The test results of the non-Gaussianity of $V$}\label{tab:Vnongauss}
\centering
\begin{tabular}{c|ccc}
\hline
\hline
$\sigma_d$ & $\langle V\rangle$ & $\sigma_V$ & $V_{\odot}$$^{\rm a}$\\
\hline
\multicolumn{4}{c}{$\kms$}\\
\hline
80 & $-16.6\pm2.5$ & $12.7\pm0.5$ &  $14.0\pm2.4$\\
110 & $-20.2\pm2.7$ & $19.9\pm1.1$& $15.2\pm2.6$\\
140 & $-32.0\pm14.1$ & $33.3\pm5.6$& $22.7\pm13.1$\\
\hline\hline
$\sigma_d$ & ${\langle V\rangle}_G$ & $\sigma_{V,G}$ & $V_{\odot,G}$$^{\rm b}$\\
\hline
\multicolumn{4}{c}{$\kms$}\\
\hline
80 & $-15.0\pm2.3$ & $11.9\pm0.4$ &$12.5\pm2.3$\\
110 &$-16.4\pm2.2$ & $16.7\pm0.5$ & $11.6\pm2.2$\\
140 & $-22.2\pm10.9$ & $23.5\pm4.4$ &$14.2\pm10.3$\\
\hline\hline
\end{tabular}
\begin{description}
\item[a] The mean and standard deviation of $V$ is calculated directly from their definitions. The $V_{\odot}$ is calculated from the direct calculated standard deviation. No Gaussianity is assumed in these derived values.\\
\item[b] The mean and sigma values are calculated from fitting a Gaussian profile. And $V_{\odot,G}$ is calculated based on the Gaussian sigma $\sigma_{V,G}$.\\
\end{description}
\end{table}

We adopt the analytical form of the distribution function defined by Equation (15) of \citet{CB94}, namely,
\begin{eqnarray}\label{DF}
&f(R,v_R,v_\phi)\sim \textrm{exp(}-{v_R^2\over{2\sigma_d^2}}e^{yv_\phi/v_c})\nonumber\\
&\times \textrm{exp}\{{1\over{2\sigma_d^2}}[v_\phi^2-v_c^2+2v_c^2ln({v_c\over{v_\phi}})]e^{yv_\phi/v_c}\},
\end{eqnarray}
where $R$ is the Galactocentric radius, $v_R$ the radial velocity, $v_\phi$ the azimuthal velocity, and $y\equiv8/R_d$. The free parameters are $R_d$, $\sigma_d$, and $v_c$. We predefine three sets of the free parameters and generate three sets of mock data via Monte Carlo simulations. Then we use Gaussian approximation to estimate the mean $v_\phi$ and $\sigma_\phi$. In all three simulations, we set $R=8$\,kpc, $v_c=220$\,$\kms$, $R_d=2.5$\,kpc. $v_R$ follows a Gaussian distribution with zero mean and sigma of $\sigma_U=\sigma_d exp(-R/R_d)$. We also add $V_{\odot}=12$\,$\kms$ in the simulations. The only varying parameter is $\sigma_d$, which is chosen as 80, 110, and 140\,$\kms$, respectively, for each simulation. We then compare with the directly estimated mean and standard deviation of $v_\phi$ without assuming a Gaussian profile. The $V_{\odot}$ is also calculated according to Equations~(\ref{eq:va}) and~(\ref{eq:vlsr}) given $L=5$\,kpc. The comparison is listed in Table~\ref{tab:Vnongauss}. This shows that $\langle V \rangle$ is underestimated by a larger amount as $\sigma_d$ increases. On the other hand, the velocity dispersion estimates from Gaussian fitting are also systematically underestimated by a few $\kms$. As a consequence, the derived $V_{\odot,G}$ from the Gaussian fitting is also slightly smaller than the true value.

Therefore, we can infer that the \V\ and $\sigma_V$ from Equation~(\ref{likelihood}) may be systematically underestimated. The same is true of the estimated $V_{\odot}$. However, it is worth noting that the velocity distribution function~(\ref{DF}) is different than that of the real Galaxy. In reality, this underestimation may be very small and can be negligible. Indeed, the validation test with GCS data shows that the reconstructed $V$ and $\sigma_V$ are only smaller than the true value within 1\,$\kms$, as shown in Table~\ref{tab_gcs}. Therefore, the $V_\odot$ estimates may be only slightly smaller than the unknown true value. Going from low to high $|z|$, the disk becomes hotter and $\sigma_d$ increases. Then, the \V\ is underestimated at higher $|z|$. However, this systematic underestimation likely occurs in all \teff\ bins and consequently cannot be the reason for the opposite trend along $|z|$ for the stars with \teff$>6000$\,K.

\subsection{The age--\teff\ relation}
\begin{figure}[htbp]
\centering
\includegraphics[scale=0.5]{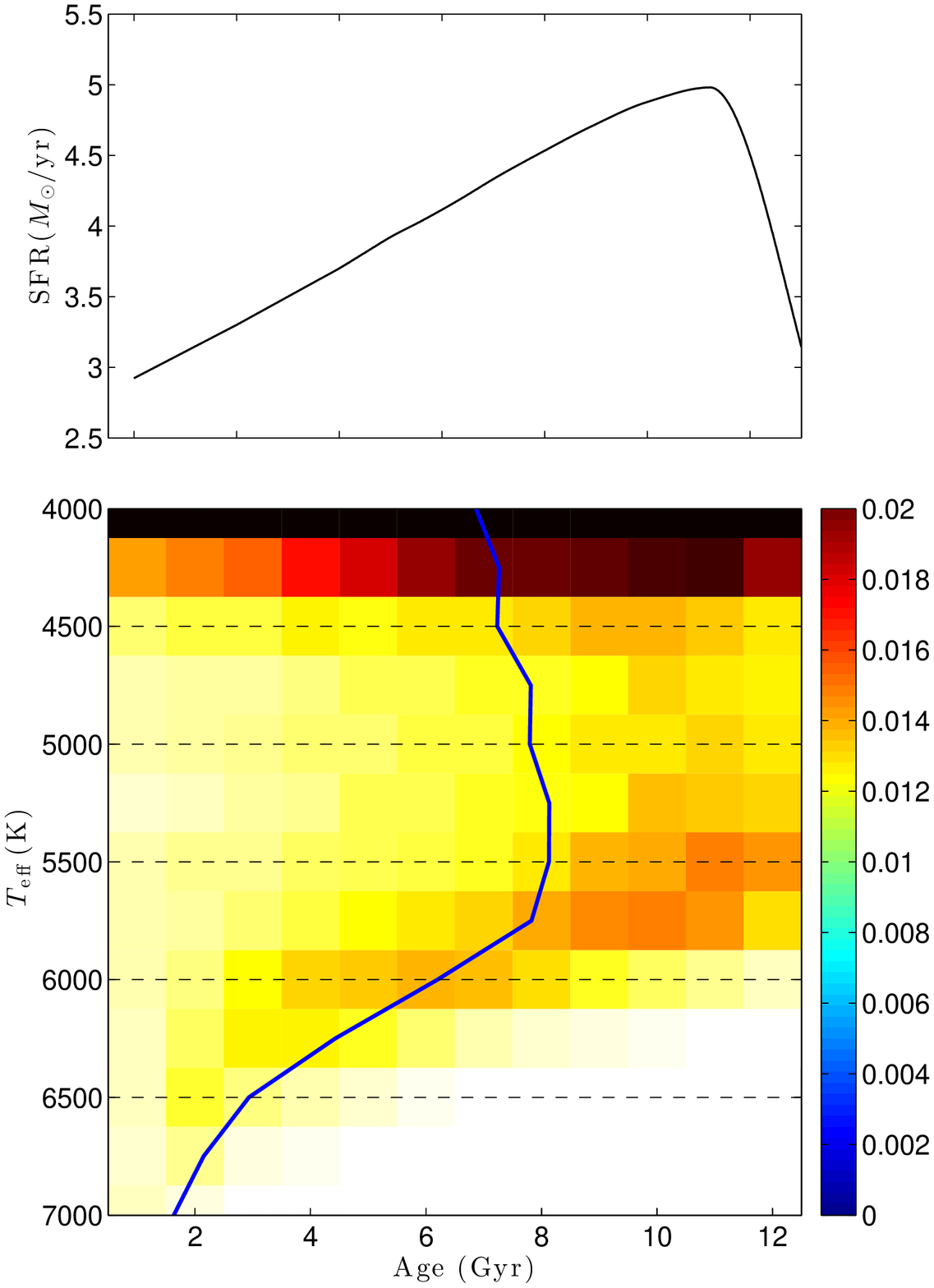}
\caption{The top panel shows the star formation history used by \citet{schoenrich09}. The bottom panel shows the stellar distribution in \teff\,vs. age plane for the main-sequence stars based on the Chabrier IMF \citep{Chabrier03} and the synthetic stellar isochrones \citep{marigo08}. The color codes the normalize stellar distribution and the blue thick line indicates the mean age as a function of \teff. The dashed lines indicate the separation of the \teff\ bins in this work.}\label{fig:teff_age}
\end{figure}

We reveal that the kinematic features for the stars with \teff$>6000$\,K are significantly different from those for the cooler stars. In this section, we first give a theoretical estimation of the age--\teff\ relation and then we discuss the possible reasons why the kinematic features are so different for the stars with \teff$>6000$\,K based on the age distribution in next section.

In order to derive the age--\teff\ relation, we first need to know the star formation history of the Milky Way. We adopt the star formation history shown in Figure~1 of \citet{schoenrich09}, which contains a star formation peak at about 11\,Gyr age and a star formation rate decreasing with time (see the top panel of Figure~\ref{fig:teff_age}). Second, we use synthetic isochrones \citep{marigo08} to select the range of \teff\ for the main-sequence stars at given age. We also adopt the initial mass function derived by \citet{Chabrier03} to assign weight for the main-sequence stars at different temperature. Finally, we set up the stellar distribution in age vs. \teff\ plane, as shown in the bottom panel of Figure~\ref{fig:teff_age}. The blue thick line shows the mean age at different \teff. It shows that for the stars with $6500<$\teff$<7000$\,K, the mean age is between 2 and 3\,Gyr and there are almost no stars with age larger than 4\,Gyr. For the stars with $6000<$\teff$<6500$\,K, the average age varies between 3 and 6\,Gyr. For the stars with \teff$<6000$\,K, the average age is around 8\,Gyr with a very broad range covering all ages. However, since the star formation rate has a peak at around 11\,Gyr, the stars with age older than 8\,Gyr dominate this region. The \teff\ of the abrupt change in age is perfectly consistent with that of the sudden change in the mean velocity and velocity dispersions shown in the left and middle panels of Figure~\ref{fig_UVW_Teff_az}. This implies that the Parenago discontinuity occurs at around 6\,Gyr. This value is about 3 Gyr earlier than the result by \citet{quillen2001}.

\subsection{The asymmetric motion of the young stars}

Similar asymmetric motion is also reported in other works. \citet{carlin2013} found both the vertical and radial velocities are not at around zero with the LAMOST data. Although the data analyzed by them are located in a larger volume than the data we use in this work, they are qualitatively consistent with the kinematic features of the young stars in our samples. \citet[][]{W13} also found similar trends in the solar vicinity with the red clump stars observed by RAVE \citep{siebert11}. Particularly, they found a faster $v_\phi$ (see their Figure 10), a negative $v_R$ (see their Figure 12), which is equivalent with a net positive $U$ in the Cartesian coordinates, and slightly positive $v_Z$ (see their Figure 13) at the position close to the Galactic mid-plane. Although the quantitative comparison is quite difficult since the spectral types of the stars in \citet{W13} are completely different and the spatial sampling is also different, the orientations and the values of the offsets in three dimensional velocity are quite similar to our results. In general, red clump stars are thought to be quite young and \citet{Girardi01} argued that the peak of the age of nearby red clump stars is only around 1\,Gyr. Then, it is not surprising that similar asymmetric motion is found in the red clump samples and the younger stars from our data set.

There are possibly three channels to explain the unusual asymmetric motions for the young stars. We discuss these three scenarios here separately.

First, if the young stars are still not completely relaxed, they may keep the peculiar motion of the molecular cloud in which they are born. Assume that the young stars are formed in a giant molecular cloud containing totally $10^3$-1$0^4$ new formed stars within a scale of $\sim$200\,pc, then the relaxation time for the group of stars is between 0.5-4\,Gyr. As shown in Figure~\ref{fig:teff_age}, for the youngest stars with \teff$>6500$\,K, even the maximum age is only $\sim2$\,Gyr. Therefore, it is possible that these young stars are still not completely relaxed and still display some kinematic features of their birth place.

Second, it is likely that the young stars are perturbed by the non-axisymmetric structure in the disk, e.g., the central rotating bar, spiral arms etc. Some studies have argued that the perturbation must affect both young and old stars \citep[e.g.][]{famaey2005}. However, in general, orbits of stars in nearly circular orbits and very close to the Galactic mid-plane can be approximated by three harmonic oscillations in azimuthal, radial, and vertical orientations. Subsequently, the relatively fixed frequencies of the oscillations make the stars more easily respond to the resonance induced by the rotating bar or spiral arms. Since young stars are mostly in near-circular orbits and concentrated in the mid-plane, they are more easily perturbed by non-axisymmetric structures than the old stars. However, the bar and spiral structures are mostly in the Galactic mid-plane and thus may not affect the vertical motion of the young stars (though see \citealt{debattista2014, faure2014} for some mechanisms by which the bar and spiral arms can vertically perturb the disk). Therefore, it is hard to explain why the young stars also show a net positive value in $W$.

Finally, the disk may have been perturbed by a merging event. Indeed, \citet{widrow2012} and \citet{yanny_gardner2013} found that the stellar vertical density shows wave-like features, which could be the density wave excited by a minor merger \citep{gomez2013}.

Certainly, none of the scenarios discussed above can be easily ruled out. Further observational and theoretical works are required to better address this challenging issue.

An interesting question raised here is why previous studies of local kinematics based on the \emph{Hipparcos} data do not show the asymmetric motion. It is quite hard to answer it. The different volume of the \emph{Hipparcos} may be one reason. More detailed investigations with the future Gaia data including accurate distance estimates as well as the 3-D velocities \citep{gaia} may be needed to complete the picture of local disk kinematics.

\subsection{The impact of systematic bias in the distance and radial velocity}\label{sect:sysbias}
\begin{figure*}[htbp]
\begin{center}
\includegraphics[height=1.3\columnwidth]{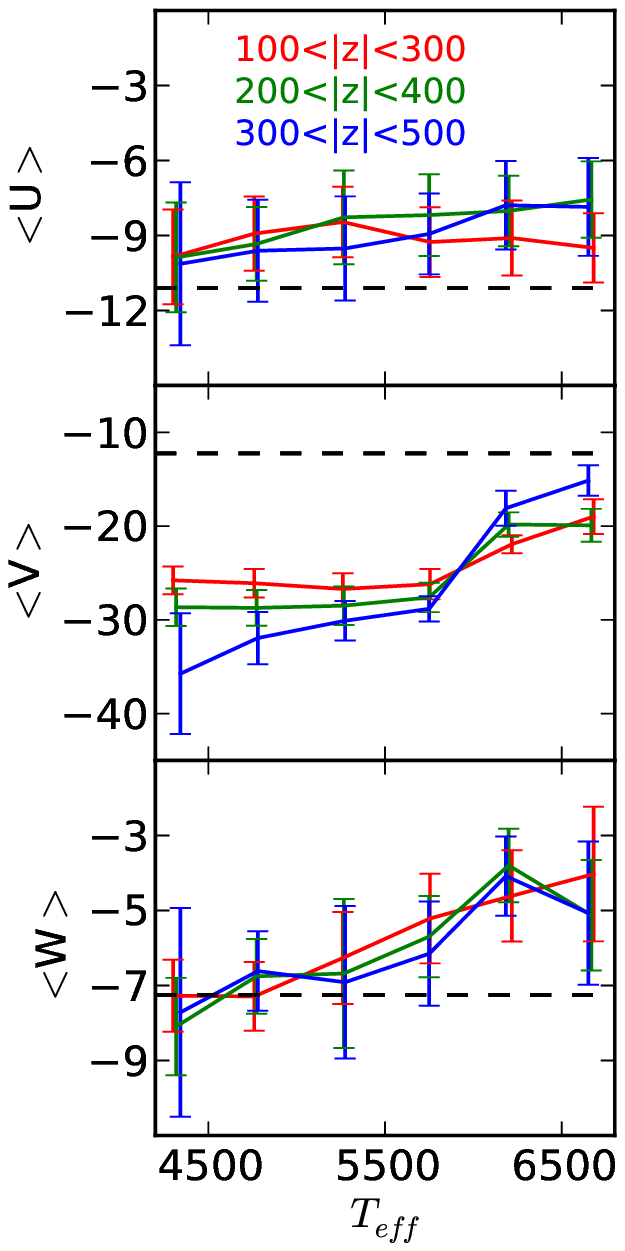}
\includegraphics[height=1.3\columnwidth]{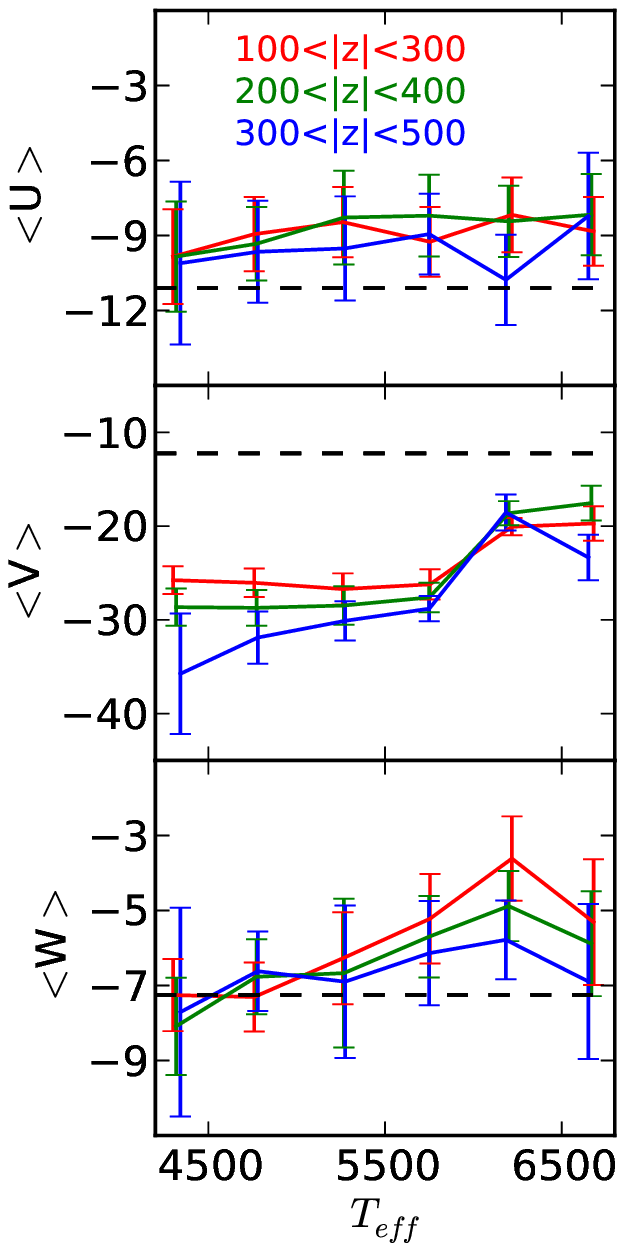}
\includegraphics[height=1.3\columnwidth]{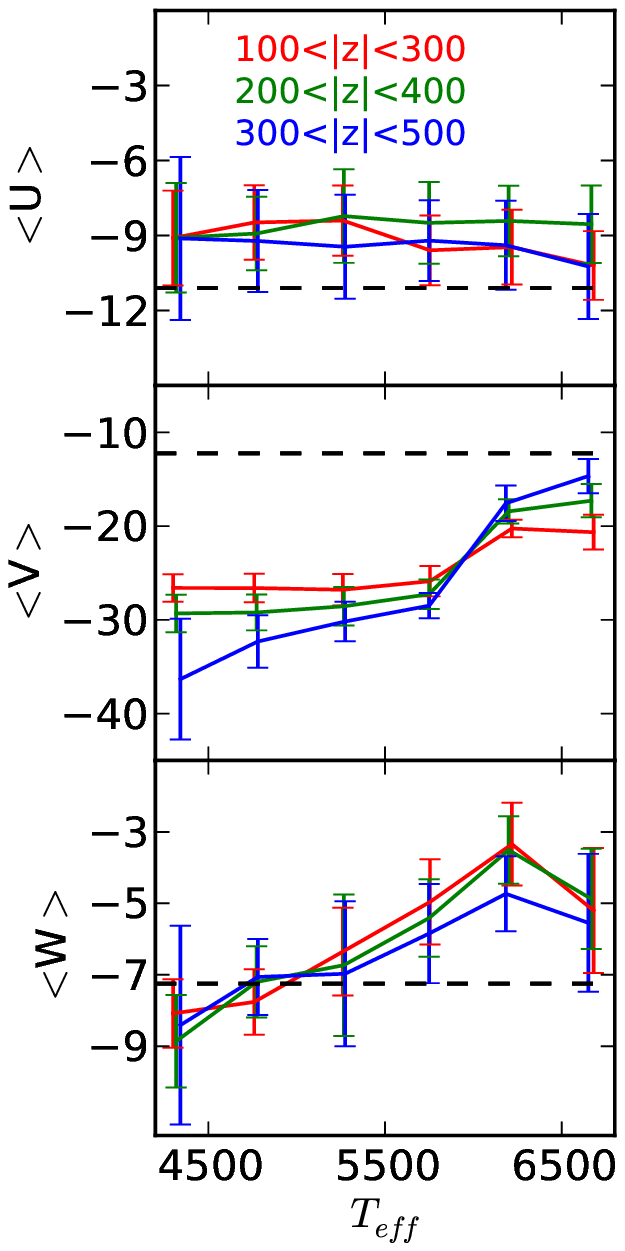}
\end{center}
\caption{The mean velocities with different systematically biased distance and radial velocity. The left panels are the results with over-estimated distance by a factor of 20$\%$, the middle panels are the results with under-estimated distance by a factor of 20$\%$ \revise{only in the bins with \teff$>$6000K and keep the distances in other bins unchanged}. And the right panels are the results with the offset \revise{varying as an anti-correlation function of \teff\ presented in Figure~\ref{fig_lm_apo}} in the radial velocity distribution.}\label{fig_20p}
\vspace{20pt}
\end{figure*}

\revise{Although Figure~\ref{fig:distresid} does not show any dependance of the distance estimates on \teff, some extreme tests, e.g., the distance is only over- or under-estimated for the warmer stars but unchanged for the cooler ones, are very helpful to verify whether the kinematic results are robust},  The left and middle panels of Figure~\ref{fig_20p} show the mean velocities derived from the distances with artificial 20\% over- and under-estimation \revise{only for the stars with \teff$>6000$\,K}, respectively. Essentially, the asymmetric motions of the young stars still exist in both \U\ and \W\ \revise{with slightly shifted values}. And the Parenago's discontinuity and turnover phenomenon in \V\ are still substantial. We also do not find any systematic bias due to the over- or under-estimation of the distance in the velocity ellipsoids.

\revise{Figure~\ref{fig_lm_apo} presents a weak anti-correlation between the offset of the radial velocities and \teff.} This drives us to further investigate whether the results are stable when the offset varies \revise{with \teff}. We then \revise{fit the offset of the radial velocity with a linear function of \teff\ and correct the radial velocities for the individual stars with the \teff-independent offset. We find} that this leads to slightly shifts in \U, \V, and \W, as shown in the right panels of Figure~\ref{fig_20p}. \revise{Although the difference in \U\ between the young and old stars is vanished, The differences in \V\ and \W\ are even stronger.} 
It may also shift the derived local motion of the Sun. For $\Usun$, it may decrease by $\sim1$\,$\kms$, and for $\Wsun$, it may increase by $\sim1$\,$\kms$. These shifts are within 1-$\sigma$ uncertainties according to Table~\ref{tab:lsr}. \revise{We also notice that \citet{gao15} found that the offset is 3.8\,$\kms$ compared the LAMOST data with PASTEL catalog~\citep{pastel}, implying that the calibration of the redial velocity may also depends on different calibrators. We did another test with the offset of 3.8\,$\kms$ and found the mean velocities shift by $\sim1$\,$\kms$ for all stars and the systematic differences between the young and old stars are unchanged. Therefore, in order to keep simplification, we adopt the correction of the radial velocity with a constant 5.7\,$\kms$ for all stars in this work. According to the tests with various corrections, we conclude that the difference in \V\ and \W\ between the young and old stars are real, while the mild difference in \U\ shown in Figure~\ref{fig_UVW_Teff_az} may be insignificant.}

\section{Conclusions}

We use the nearby FGK type main-sequence stars selected from LAMOST DR1 data to estimate the three dimensional velocities and velocity ellipsoids from line-of-sight velocities alone. It has been known that the velocity de-projection technique can introduce systematic bias due to the spatial variation of the velocity dispersions. Moreover, we find that the uneven spatial sampling can also affect the velocity dispersion as well. In order to derive the corrected velocity ellipsoid from the de-projection method, we calibrate it using a set of transform matrices estimated from simulations. The calibration works well in the simulations and then is applied to the observed data.

We associate the derived velocities and their ellipsoids with the effective temperature and $|z|$ and reveal that the asymmetric motions of the stars in the solar neighborhood reported by previous works are mainly seen among warm stars with \teff$>6000$\,K, which are very young with average age of less than 4\,Gyr. These young stars rotate faster in larger $|z|$ than in smaller $|z|$. \revise{Meanwhile, they move up toward the north Galactic pole by about 3\,$\kms$ and probably radially inward to the Galactic center by a few $\kms$}. The nature of the asymmetric motion is still not clear.  With the older (cool) stars, we give estimates of the solar motion with respect to the LSR. We obtain ($U_\odot$, $V_\odot$, $W_\odot$)$=$(9.58$\pm$2.39, 10.75$\pm$1.96, 7.01$\pm$1.67)\,$\kms$.

We also derive the velocity ellipsoids and find that the young stars have significantly smaller dispersions than the older stars. The vertical gradient of the velocity dispersions is larger in $\sigma_V$ and $\sigma_W$ than in $\sigma_U$. On the other hand, $\sigma_U$ shows clear correlation with \teff\ but the other two dispersions do not. We confirm that the Parenago discontinuity occurs at about 6\,Gyr by comparing the velocity dispersions with a simple star formation model.

The derived velocity ellipsoids in this work are essentially consistent with those in DB98, S12, and B14. The velocity de-projection method may still slightly suffer from the distortion in the derived cross terms even after calibration. Therefore, more accurate estimations about the orientation of the velocity ellipsoid should be done with three dimensional velocities.



\acknowledgements

\revise{We thank the anonymous referee for his/her very helpful comments.} This work is supported by the Strategic Priority Research Program
``The Emergence of Cosmological Structures" of the Chinese Academy of Sciences, Grant No. XDB09000000 and the National Key Basic
Research Program of China 2014CB845700. THJ acknowledges the National Natural Science Foundation of China (NSFC) under grants U1231123, U1331202, U1331113, 11303020, and the LAMOST Fellowship. CL acknowledges the NSFC under grants 11373032, 11333003, and U1231119. JLC is supported by US National Science Foundation grants AST 09-37523 and AST 14-09421. XLC thanks the support by the MoST 863 program grant 2012AA121701, the NSFC grants 11373030.
The Guoshoujing Telescope (Large Sky Area Multi-Object Fiber Spectroscopic Telescope, LAMOST) is a National Major Scientific Project built by the Chinese Academy of Sciences. Funding for the project
has been provided by the National Development and Reform Commission. LAMOST is operated and managed by the National Astronomical Observatories, Chinese Academy of Sciences.

{\it Facilities:} \facility{LAMOST}.


\clearpage
\begin{turnpage}
\begin{table}
\caption{The derived velocities and the velocity ellipsoids from the LAMOST data}\label{tab_sigma2b}
\begin{center}
\scriptsize
\begin{tabular}{c c c c c c c c c c c c c c c}
\hline
\hline
\hline
Teff&$\langle N\rangle$& $\langle U\rangle$& $\langle V\rangle$& $\langle W\rangle$& $\sigma_{U}$& $\sigma_{V}$& $\sigma_{W}$& $\sigma_{UV}$& $\sigma_{UW}$& $\sigma_{VW}$ & $\sigma_{U}/\sigma_{V}$& $\sigma_{U}/\sigma_{W}$ & $l_v$& $\alpha$\\
K&&\multicolumn{9}{c}{$\kms$}&&&degree&degree\\
\hline
\hline
\multicolumn{15}{c}{100$<|z|<$300}\\
\hline
4250&13229&-10.41$\pm$1.90&-26.09$\pm$1.48&-7.26$\pm$0.96&41.25$\pm$1.84&27.16$\pm$2.11&22.54$\pm$1.10&-6.51$\pm$10.23&-4.21$\pm$7.61&-5.00$\pm$15.45&1.52$\pm$0.09&1.83$\pm$0.07&-0.08$\pm$7.41&-0.85$\pm$3.92\\
4750&19593&-8.74$\pm$1.49&-25.96$\pm$1.52&-7.17$\pm$0.92&40.77$\pm$1.30&27.12$\pm$2.14&23.43$\pm$0.91&-5.96$\pm$9.15&-5.40$\pm$8.12&3.57$\pm$7.98&1.50$\pm$0.09&1.74$\pm$0.05&-2.17$\pm$8.12&-1.50$\pm$4.05\\
5250&25805&-8.76$\pm$1.41&-26.53$\pm$1.68&-6.35$\pm$1.23&40.79$\pm$1.60&27.10$\pm$1.55&23.17$\pm$1.26&-5.16$\pm$6.28&-4.83$\pm$6.01&3.47$\pm$8.18&1.51$\pm$0.07&1.76$\pm$0.07&-1.63$\pm$5.90&-1.19$\pm$8.58\\
5750&33453&-8.86$\pm$1.40&-26.45$\pm$1.62&-5.35$\pm$1.19&39.86$\pm$1.36&26.68$\pm$2.00&21.80$\pm$1.04&-5.29$\pm$9.62&-9.70$\pm$7.09&5.63$\pm$5.14&1.49$\pm$0.08&1.83$\pm$0.06&-1.80$\pm$8.98&-4.80$\pm$3.86\\
6250&23565&-8.92$\pm$1.50&-21.44$\pm$0.93&-3.64$\pm$1.16&32.97$\pm$1.17&20.21$\pm$1.99&17.36$\pm$0.95&2.31$\pm$6.00&-2.70$\pm$12.51&-4.41$\pm$6.65&1.63$\pm$0.10&1.90$\pm$0.07&0.45$\pm$6.62&-0.53$\pm$4.25\\
6750&6739 &-9.07$\pm$1.37&-23.17$\pm$1.85&-4.57$\pm$1.75&26.05$\pm$1.05&19.68$\pm$1.97&16.74$\pm$1.86&-1.18$\pm$9.26&1.79$\pm$3.47&1.05$\pm$7.94&1.32$\pm$0.11&1.56$\pm$0.12&-0.27$\pm$9.63&0.46$\pm$6.22\\
\hline
\hline
\multicolumn{15}{c}{200$<|z|<$400}\\
\hline
4250&7600&-12.22$\pm$2.19&-28.47$\pm$2.00&-7.88$\pm$1.30&40.79$\pm$2.19&31.53$\pm$2.86&24.83$\pm$1.44&-2.11$\pm$12.69&-11.21$\pm$9.35&-2.24$\pm$8.92&1.29$\pm$0.11&1.64$\pm$0.08&-0.68$\pm$10.41&-6.75$\pm$4.42\\
4750&16639&-8.98$\pm$1.47&-29.05$\pm$1.91&-6.90$\pm$1.00&42.34$\pm$2.07&28.69$\pm$2.23&25.62$\pm$1.09&-9.78$\pm$11.28&-9.71$\pm$3.75&2.56$\pm$11.48&1.48$\pm$0.09&1.65$\pm$0.06&-5.56$\pm$10.81&-4.71$\pm$3.95\\
5250&27523&-8.20$\pm$1.87&-28.68$\pm$2.05&-6.77$\pm$1.98&42.73$\pm$1.50&28.40$\pm$2.43&24.62$\pm$1.05&3.65$\pm$4.46&-4.40$\pm$5.92&1.80$\pm$7.40&1.50$\pm$0.09&1.74$\pm$0.06&0.76$\pm$5.18&-0.91$\pm$4.56\\
5750&36303&-8.08$\pm$1.63&-27.97$\pm$1.57&-5.73$\pm$1.09&42.19$\pm$1.47&30.13$\pm$1.89&24.61$\pm$1.33&8.79$\pm$9.63&-7.39$\pm$4.78&-2.15$\pm$7.90&1.40$\pm$0.07&1.71$\pm$0.06&4.99$\pm$7.87&-2.66$\pm$5.12\\
6250&18621&-7.39$\pm$1.41&-19.89$\pm$1.29&-4.01$\pm$0.95&35.02$\pm$1.50&20.92$\pm$2.14&20.29$\pm$1.09&8.12$\pm$3.20&-5.66$\pm$6.71&-0.44$\pm$5.57&1.67$\pm$0.11&1.73$\pm$0.07&4.73$\pm$6.51&-2.25$\pm$4.76\\
6750&3436 &-6.49$\pm$1.54&-20.47$\pm$1.77&-5.28$\pm$1.40&27.27$\pm$1.39&20.42$\pm$1.77&16.94$\pm$1.34&-4.24$\pm$10.13&3.21$\pm$4.88&-1.41$\pm$7.43&1.34$\pm$0.10&1.61$\pm$0.09&-3.16$\pm$12.35&1.29$\pm$7.38\\
\hline
\hline
\multicolumn{15}{c}{300$<|z|<$500}\\
\hline
4250&3797&-10.58$\pm$3.25&-35.84$\pm$6.44&-7.49$\pm$2.78&41.21$\pm$10.82&38.06$\pm$5.93&26.67$\pm$6.46&9.62$\pm$17.82&-11.10$\pm$10.02&0.34$\pm$9.78&1.08$\pm$0.31&1.55$\pm$0.36&27.20$\pm$14.35&-7.01$\pm$7.32\\
4750&11443&-9.37$\pm$2.04&-32.45$\pm$2.79&-6.33$\pm$1.06&42.85$\pm$2.12&31.70$\pm$2.44&27.20$\pm$5.43&5.96$\pm$9.46&-7.46$\pm$4.18&0.07$\pm$9.94&1.35$\pm$0.09&1.58$\pm$0.21&1.80$\pm$13.30&-2.90$\pm$3.53\\
5250&22503&-9.03$\pm$2.08&-30.24$\pm$2.10&-6.98$\pm$2.03&43.33$\pm$2.80&29.96$\pm$2.79&25.78$\pm$1.56&8.68$\pm$8.89&-6.24$\pm$11.04&-4.10$\pm$9.81&1.45$\pm$0.11&1.68$\pm$0.09&4.47$\pm$8.44&-1.84$\pm$6.28\\
5750&33595&-8.92$\pm$1.62&-29.11$\pm$1.35&-5.85$\pm$1.39&42.68$\pm$2.17&32.81$\pm$2.67&26.71$\pm$1.39&14.13$\pm$7.54&-4.91$\pm$2.45&-2.65$\pm$8.72&1.30$\pm$0.10&1.60$\pm$0.07&14.10$\pm$8.03&-1.25$\pm$4.87\\
6250&13986&-8.40$\pm$1.78&-18.99$\pm$1.88&-4.64$\pm$1.05&37.32$\pm$2.01&26.09$\pm$2.84&21.81$\pm$1.32&11.13$\pm$5.17&-5.26$\pm$3.04&-3.20$\pm$7.56&1.43$\pm$0.12&1.71$\pm$0.08&9.60$\pm$8.41&-1.73$\pm$5.93\\
6750&1566 &-7.24$\pm$2.09&-17.39$\pm$1.81&-6.62$\pm$1.91&30.64$\pm$7.05&20.84$\pm$7.12&22.42$\pm$6.33&-7.45$\pm$9.12&7.57$\pm$11.32&2.71$\pm$5.20&1.47$\pm$0.41&1.37$\pm$0.36&-6.71$\pm$12.79&7.36$\pm$10.43\\
\hline
\hline
\hline
\end{tabular}
\end{center}
\end{table}
\end{turnpage}
\clearpage


\begin{thebibliography}{}

\bibitem[{{Ahn} et al. (2014)}]{ahn14}Ahn, C.~P., Alexandroff, R., Allende Prieto, C., et al. 2014, \apjs, 211, 17

\bibitem[{{Antoja} {et~al.}(2012){Antoja}}]{A12}{Antoja}, T., et al. 2012, MNRAS, 426,
  L1-L5

\bibitem[{{Antoja et al.} (2011) {Antoja}}]{antoja2011}Antoja, T., Figueras, F., Romero-G\'omez., M. et al. 2011, \mnras, 418, 1423
\bibitem[{{Binney}(2010)}]{B10}
{Binney}, J.~J. 2010, \mnras, 401, 2318

\bibitem[{{Binney} \& {Tremaine}(2008) {Binney} {Tremaine}}]{BT08}
{Binney}, J., {Tremaine}, S. 2008, {Galactic Dynamics: Second Edition}
  Princeton University Press
  
\bibitem[{{Binney et al.}(2014)}]{B14}
{Binney}, J.~J., {Burnett}, B., {Kordopatis}, G., et al. 2014, \mnras, 439, 1231B (B14)
\bibitem[Bovy et al.(2014)]{bovy14} Bovy, J., Nidever, D.~L., 
Rix, H.-W., et al.\ 2014, \apj, 790, 127 

\bibitem[{{B\"udenbender, van de Ven \& Watkins}(2014)}]{BvW14}B\"udenbender, A., van de Ven, G., Watkins, L.~L. arXiv:1407.4808

\bibitem[{{Carlin et al.} (2013)}]{carlin2013} {Carlin}, J. L., {DeLaunay}, J., {Newberg}, H. J., {et~al.} 2013, \apjl, 777, 5

\bibitem[{{Carlin et al.} (2015) {Carlin}}]{carlin15} {Carlin}, J. L., Luy, C., Newberg, H. J. {et~al.}, 2015, \aj, in press, arXiv:1505.05521 

\bibitem[{{Casetti} {et~al.} (2011) {Casetti} {Girard} {Korchagin}}]{Casetti11} {Casetti-Dinescu}, D. I., {Girard}, T. M., {et~al.} 2011, \apj, 728, 7

\bibitem[{{Chabrier} (2003) {Chabrier}}]{Chabrier03} {Chabrier}, G. 2003, \pasp, 115, 763

\bibitem[{{Cuddeford} \& {Binney}(1994) {Cuddeford} {Binney}}]{CB94}
{Cuddeford}, P., \& {Binney}, J. 2012, \mnras, 266, 273

\bibitem[{{Cui} {et~al.}(2012) {Cui} {Zhao} {Chu}}]{Cui2012}
{Cui}, X.~Q., {Zhao}, Y.~H., {Chu}, Y.~Q., {et~al.} 2012, RAA, 12, 1197

\bibitem[Debattista(2014)]{debattista2014} Debattista, V.~P.\ 2014,
\mnras, 443, L1

\bibitem[{{Dehnen} (1998) {Dehnen}}]{dehnen1998}{Dehnen}, W. 1998, \aj, 115, 2384

\bibitem[{{Dehnen} \& {Binney}(1998)}]{DB98}{Dehnen}, W., {Binney}, J.~J. 1998, \mnras, 298, 387 (DB98)

\bibitem[{{Dehnen} (2000) {Dehnen}}]{dehnen2000} {Dehnen}, W. 2000, \aj, 119, 800

\bibitem[{{Deng et al.} (2012) {Deng}}]{Deng2012}
  {Deng}, L. C., {Newberg}, H. J., {Liu}, C., et al. 2012, RAA, 12, 735

\bibitem[{{Famaey et al.} (2005) {Famaey}}]{famaey2005}
 {Famaey}, B., {Jorissen}, A., {Luri}, X., et al. 2005, A\&A, 430, 165

\bibitem[Faure et al.(2014)]{faure2014} Faure, C., Siebert, A.,
\& Famaey, B.\ 2014, \mnras, 440, 2564

\bibitem[{{Foreman-Mackey} {et~al.}(2012) {Foreman-Mackey}}]{Foreman2012}{Foreman-Mackey}, D., {Hogg}, D. W., {Lang}, D., {Goodman}, J. 2012, \apj, 752, 147

\bibitem[{{Fuchs} {et~al.}(2009){Fuchs} {Dettbarn} {Rix}}]{F09}
 {Fuchs}, B., {Dettbarn}, C., {Rix}, H. W., {et~al.} 2009, \aj, 137, 4149

\bibitem[{{Fux} (2001) {Fux}}]{fux2001}
  {Fux}, R. 2001, A\&A, 373, 511

\bibitem[{{Gao} et al. (2014)}]{gao14}Gao, S., Liu, C., Zhang, X. B., et al. 2014, \apjl, 788, 37

\bibitem[Gao et al.(2015)]{gao15}Gao, H, Zhang, H., Xiang, M. et al., 2015, submitted to RAA

\bibitem[{{Girardi} \& {Salaris} (2001) {Girardi}}]{Girardi01}
 {Girardi}, L. \& {Salaris}, M., 2001, \mnras, 323, 109
\bibitem[Gilmore et al.(2012)]{gaiaeso} Gilmore, G., Randich, S., Asplund, M., et al.\ 2012, The Messenger, 147, 25 

\bibitem[{{G\'omez et al. } (2013) {G\'omez}}]{gomez2013}
 {G\'omez}, F. A., {Minchev}, I., {O'Shea}, B. W., et al. 2013, \mnras, 429,159

\bibitem[{{Goodman} \& {Weare}(2010) {Goodman} {Weare}}]{Goodman2010}{Goodman}, J., \& {Weare}, J. 2010, Comm. App. Math. Comp. Sci., 5(1), 65

\bibitem[Huang et al.(2015)]{huang15} Huang, Y., Liu, X.-W., Yuan, H.-B., et al.\ 2015, \mnras, 449, 162 

\bibitem[{{Holmberg}, {Nordstr\"om} \& {Anderson} (2007) {Holmberg} {Nordstr\"om} {Anderson}}]{holmberg2007}{Holmberg}, J., {Nordstr\"om}, B., \& {Anderson}, J. 2007, A\&A, 501, 941


\bibitem[{{Jenkins}(1992)}]{Jenkins1992}
{Jenkins}, A. 1992, \mnras, 257, 620

\bibitem[{{Juri{\'c}} {et~al.}(2008){Juri{\'c}} {Ivezi{\'c}} {Brooks} \& {et~al.}}]{Juricetal2008_short}
 {Juri{\'c}}, M., {Ivezi{\'c}}, {\v Z}., {Brooks}, A., {et~al.} 2008, \apj, 673, 864

\bibitem[Kordopatis et al.(2013)]{rave} Kordopatis, G., 
Gilmore, G., Steinmetz, M., et al.\ 2013, \aj, 146, 134





\bibitem[{{Marigo} et al. (2008)}]{marigo08}Marigo, P., Girardi, L., Bressan, A., et al. 2008, A\&A, 482,883

\bibitem[{{McMillan} \& {Binney}(2009)}]{MB09}
{McMillan}, P.~J., {Binney}, J.~J. 2009, \mnras, 400, 103



\bibitem[{{Nordstr\"om} {et~al.}(2004) {Nordstr\"om}}]{N04}
{Nordstr\"om}, B., {et~al.} 2004, A\&A, 418, 989

\bibitem[{{Parenago} (1950) {Parenago}}]{parenago1950}
{Parenago}, P.~P. 1950, AZh, 27, 150

\bibitem[{{Perryman} et al. (2001)}]{gaia}Perryman, M. A. C., de Boer, K. S., Gilmore, G. et al., 2001, A\&A, 369, 339
\bibitem[{{Quillen \& Garnett} (2001) {Quillen}}]{quillen2001}
 {Quillen}, A. C. \& {Garnett}, D. R. 2001, in Galaxy Disks and Disk Galaxies, ed. G. Jose, Funes, S. J., \& E. M. Corsini, eds. ASP Conf. Ser., 230, 87
\bibitem[Recio-Blanco et al.(2014)]{racio14} Recio-Blanco, A., de Laverny, P., Kordopatis, G., et al.\ 2014, \aap, 567, A5 


\bibitem[{{Roman}(1950)}]{Roman1950}
{Roman}, N. G. 1950, \apj, 112, 554

\bibitem[{{Roman}(1952)}]{Roman1952}
{Roman}, N. G. 1952, \apj, 116, 122

\bibitem[R\"oser, Demleitner \& Schilbach(2010)]{ppmxl}R\"{o}ser, S., Demleitner, M., \& Schilbach, E. 2010, \aj, 139, 244
\bibitem[Soubiran et al.(2010)]{pastel}Soubiran, C., Le Campion, J.-F., Cayrel de Strobel, G., \& Caillo, A., 2010, \aap, 515, 111
Wu, Y.,
\bibitem[{{Sch\"onrich} \& {Binney} (2009)}]{schoenrich09}Sch\"onrich, R. \& Binney, J., 2009, \mnras, 396, 203


\bibitem[{{Sch\"onrich}, {Binney} \& {Dehnen} (2010)}]{SBD10}
{Sch\"onrich}, R., {Binney}, J., \& {Dehnen}, W., 2010, \mnras, 403, 1829

\bibitem[{{Sch\"onrich} \& {Binney} (2012)}]{schoenrich12}Sch\"onrich, R. \& Binney, J., 2012, \mnras, 419, 1546

\bibitem[{{Schwarzschild}(1908)}]{Schwarzschild1908}
{Schwarzschild}, K. 1908, Nachr. Kgl. Ges. d. Wissenschaften, G\"ottingen, p. 191

\bibitem[{{Sekiguchi} \& {Fukugita}(2000) {Sekiguchi}{Fukugita}}]{SF00}
{Sekiguchi}, M., {Fukugita}, M. 2000, \aj, 120, 1072

\bibitem[{{Sharma} et al. (2014)}]{Sharma14}
{Sharma}, S., {Bland-Hawthorn}, J., {Binney}, J., et al. 2014, \apj, 793, 51S

\bibitem[{{Siebert} et al. (2008)}]{siebert08} Siebert A., Bienaym\ 'e, O., Binney, J., et al., 2008, \mnras, 391, 793

\bibitem[{{Siebert} et al. (2011)}]{siebert11}Siebert, A., Williams, M. E. K., Siviero, A. {et al.} 2011, \aj, 141, 187


\bibitem[{{Smith, Whiteoak \& Evans} (2012) {Smith} {Whiteoak} {Evans}}]{smith2012}
 {Smith}, M. C., {Whiteoak}, S. H. \& {Evans}, N. W., 2012, \apj, 746, 181(S12)

\bibitem[{{Sofue} {et~al.}(2009)}]{SHO09}
{Sofue}, Y., {Honma}, M., \& {Omodaka}, T. 2009, PASJ, 61, 227

\bibitem[{{van der Kruit} \& {Freeman} (2011)}]{kf11}van der Kruit, P.~C. \& Freeman, K.~C., 2011, \araa, 49, 301

\bibitem[{{Widrow} {et~al.} (2012) {Widrow}}]{widrow2012}
 {Widrow}, L. M., {Gardner}, S., {Yanny}, B., et al. 2012, \apjl, 750, 41

\bibitem[{{Williams} {et~al.}(2013){Williams}}]{W13}
 {Williams}, M. E. K., et al. 2013, \mnras, 436, 101

\bibitem[{{Xia} {et~al.} (2014) {Xia}}]{xia14}{Xia}, Q. R, {Liu}, C., {Mao}, S., {et~al.} 2014, MNRAS in press, arXiv1412.4770

\bibitem[Yanny \& Gardner(2013)]{yanny_gardner2013} Yanny, B., \& Gardner, S.\, 2013, \apj, 777, 91

\bibitem[Zacharias et al.(2013)]{ucac4} Zacharias, N., Finch, 
C.~T., Girard, T.~M., et al.\ 2013, \aj, 145, 44

\bibitem[{{Zhao} {et~al.}(2012) {Zhao} {Zhao} {Chu}}]{Zhao2012}
{Zhao}, G., {Zhao}, Y.~H., {Chu}, Y.~Q., {et~al.} 2012, RAA, 12, 723

\bibitem[{{Zhao, Zhao \& Chen} (2009) {Zhao} {Zhao} {Chen}}]{zhao2009}{Zhao}, J. K., {Zhao}, G., \& {Chen}, Y. Q., 2009, \apjl, 692, 113

\end{thebibliography}
\end{document}